\documentclass[reprint,
 amsmath, amssymb,
 aps, floatfix, showkeys
]{revtex4-2}

\usepackage{graphicx}
\usepackage{dcolumn}
\usepackage{bm}
\usepackage[colorlinks, linkcolor = magenta, urlcolor = blue]{hyperref}
\usepackage{braket}
\usepackage{physics}
\usepackage{amsmath}
\usepackage{array}
\usepackage{float}
\usepackage{multirow}

\begin{document}
\title{Protecting quantum correlations of negative quantum states using weak measurement under non-Markovian noise}
\author{Jai Lalita}
\email{jai.1@iitj.ac.in}
\author{Subhashish Banerjee}
\email{subhashish@iitj.ac.in }
\affiliation{Indian Institute of Technology, Jodhpur-342030, India}

\date{\today}

\begin{abstract}
The weak measurement (WM) and quantum measurement reversal (QMR) are crucial in protecting the collapse of quantum states. The idea of WM and QMR has recently been used to protect and enhance quantum correlations and universal quantum teleportation (UQT) protocols. Here, we study the quantum correlations, maximal fidelity, and fidelity deviation of the two-qubit negative quantum states developed using discrete Wigner functions with(without) WM and QMR. To take into account the effect of a noisy environment, we evolve the states via non-Markovian amplitude damping and random telegraph noise quantum channels. To benchmark the performance of negative quantum states, we calculate their success probability. We compare our results with the two-qubit maximally entangled Bell state. Interestingly, we observe that some negative quantum states perform better with WM and QMR than the Bell state for different cases under evolution via noisy quantum channels.
\end{abstract}

\keywords{Negative quantum states, weak measurement, and quantum measurement reversal.}

\maketitle
\section{\label{sec:level1}Introduction\protect}
Quantum correlations play a vital role in many quantum information applications, including quantum teleportation \cite{bennett1993teleporting, bouwmeester1997experimental, adhikari2012operational}, quantum computing \cite{nielsen2010quantum}, quantum cryptography \cite{masanes2011secure}, and quantum metrology \cite{giovannetti2011advances, thapliyal2017quantum}. A central driving element for realizing quantum teleportation (QT) was its prospective application in the technological advancement of communication \cite{gisin2007quantum, jin2010experimental, zeilinger2018quantum}. The approach presented in \cite{jozsa1993teleporting} was shown to be feasible by the experimental demonstrations on a single arbitrary qubit state, which involved the destruction and reconstruction of the quantum state using classical information and the non-local correlation of the EPR (Einstein-Podolsky-Rosen) channel \cite{bennett1993teleporting, boschi1998experimental}. For QT, the average fidelity is the standard figure of merit \cite{horodecki1996teleportation, badziag2000local}. It is the average overlap or closeness between Bob's input state and Charlie's received output state, calculated by averaging across all potential input states. In reality, all quantum systems are open and constantly interact with their immediate environment, which causes the degradation of non-local correlations \cite{wootters1998entanglement, horodecki2009quantum, ollivier2001quantum, henderson2001classical, bennett1999quantum, schrodinger1935discussion, schrodinger1936probability, brunner2014bell, fan2022quantum, costa2016quantification, Chakrabarty2010study, luo2008quantum, ramkarthik2020quantum} and, hence, average fidelity. 
Moreover, that introduces variance in average fidelity values for various possible input states. So, to further explore the effect of noise on QT, we employ fidelity deviation in addition to average fidelity \cite{bang2018fidelity}. The highest achievable average fidelity value across all feasible local unitary operations in the standard teleportation protocol is the maximal fidelity \cite{horodecki1996teleportation}. The goal is to minimize fidelity deviation while maintaining the highest feasible average fidelity \cite{ghosal2020optimal}. So, for QT, a suitable state is any two-qubit entangled state where the maximal fidelity is strictly greater than the classical constraint \cite{horodecki1996teleportation, horodecki1999general}. On the contrary, if and only if a state exhibits zero fidelity deviation, it is considered useful for universal quantum teleportation (UQT) \cite{ghosal2021characterizing}.

The fundamental basis for examining how an environment influences a quantum system is provided by the theory of open quantum systems \cite{breuer2002theory, nielsen2010quantum, banerjee2018open, weiss2012quantum}. When there is a clear distinction between the system and environment time scales, the dynamics of an open quantum system can be described by the Markovian approximation \cite{breuer2002theory}. If this is not so, we enter into the non-Markovian regime \cite{rivas2014quantum, li2018concepts, breuer2016colloquium, daffer2004depolarizing, kumar2018non, utagi2020temporal, tiwari2023impact}. The back-flow of information from the environment into the system and deviation from the CP-divisibility criterion are well-known indicators of non-Markovian behavior \cite{breuer2009measure, chruscinski2011measures, rivas2010entanglement}. Moreover, the ideas of open quantum systems have numerous applications discussed in \cite{caldeira1983quantum, grabert1988quantum, hu1994quantum, banerjee2003general, plenio2008dephasing, banerjee2017characterization, de2017dynamics, rivas2014quantum, li2018concepts, breuer2016colloquium, daffer2004depolarizing}.

Thus, to advance quantum communication and computation, tackling the degradation of quantum correlations is crucial. For this, quantum measurements, $\textit{i.e.}$, weak measurement (WM), and quantum measurement reversal (QMR) have been used. The WM extracts information from the system without causing it to collapse into an eigenstate, unlike the conventional von Neumann measurement \cite{aharonov1988result, oreshkov2005weak, korotkov2006undoing, katz2008reversal, kim2009reversing, korotkov2010decoherence, kim2012protecting, dressel2014colloquium, lahiri2021exploring, sabale2023towards}. A proper QMR can reconstruct the state with a certain probability, also called the success probability \cite{pramanik2013improving, he2020enhancing}. Therefore, it has been shown that the WM and QMR can enhance and protect the quantum correlations of the qubit and qutrit quantum systems from the effects of noise, in particular, non-Markovian noise \cite{korotkov2010decoherence, kim2012protecting, xiao2013protecting, sun2017recovering}. Further, these quantum measurements can also protect and elevate UQT requirements for two-qubit quantum systems \cite{horodecki1996teleportation, badziag2000local, bang2018fidelity, ghosal2020optimal}. Additionally, using photonic and superconducting quantum systems, WM and QMR can be implemented experimentally \cite{monroe2021weak, katz2008reversal, kim2009reversing}. 

In this paper, in order to find suitable two-qubit states, apart from the Bell states, for universal quantum teleportation, we study the impact of weak measurement (WM) and quantum measurement reversal (QMR) \cite{korotkov2006undoing, katz2008reversal, kim2009reversing, korotkov2010decoherence, kim2012protecting, sabale2023towards} on the quantum correlations \cite{wootters1998entanglement, horodecki2009quantum, ollivier2001quantum, henderson2001classical, bennett1999quantum, schrodinger1935discussion, schrodinger1936probability, brunner2014bell, fan2022quantum, costa2016quantification} and UQT requirements \cite{horodecki1996teleportation, badziag2000local, bang2018fidelity, ghosal2020optimal} of the negative quantum states of two-qubit systems proposed in \cite{lalita2023harnessing}.  The influence of both non-Markovian non-unital (specified by non-Markovian amplitude damping (AD)) and unital  (specified by non-Markovian random telegraph noise (RTN)) channels are taken into consideration. Precisely, negative quantum states are non-classical states corresponding to the normalized eigenvectors of the negative eigenvalues of the phase-space point operators of discrete Wigner functions, reviewed in Sec. \ref{DWF&NQS}. We also compare the variations of quantum correlations, maximal fidelity, fidelity deviation, and success probability of two-qubit negative quantum states with that of the maximally correlated state, i.e., the Bell state, in the presence and absence of WM and QMR under the above-mentioned quantum channels.

The outline of the paper is as follows. In Sec. \ref{Model}, we briefly discuss the discrete Wigner functions and negative quantum states. This is followed by noisy quantum channels and a tentative physical model, where the notions of weak measurement and quantum measurement reversal are introduced. Section \ref{protectingQCs} presents a short review of quantum correlations, maximal fidelity, fidelity deviation, and universal quantum teleportation and studies their behavior for two-qubit negative quantum states and the Bell states under non-Markovian non-unital (amplitude damping) and unital (random telegraph noise) channels with(without) weak measurement and quantum measurement reversal. Our results are discussed in Sec. \ref{result&discussion}, followed by the conclusions in Sec. \ref{conclusion}.

\section{\label{Model} Model}
This section briefly reviews negative quantum states \cite{wootters2004picturing, gibbons2004discrete, lalita2023harnessing}, and the non-Markovian noisy quantum channels. Additionally, a physical model for protecting quantum correlations and universal quantum teleportation protocols of two-qubit quantum states using weak measurement and quantum measurement reversal in the non-Markovian environment is also canvased.

\subsection{\label{DWF&NQS}Negative quantum states}
To find the negative quantum states, we emphasize the class of discrete Wigner functions for power-of-prime dimensions, say $d$. This class defines discrete Wigner functions by associating lines in discrete phase space to projectors belonging to a predetermined set of mutually unbiased bases (MUBs) \cite{lidl1994introduction, wootters1989MUB, Lawrence2002MUB, bandyopadhyay2002MUB, durt2010mutually}. In this formulation, the discrete Wigner functions are uniquely defined as the sum of the Wigner function elements corresponding to each line equal to the probability of projecting onto the bases vector mapped with that line \cite{wootters2004picturing, gibbons2004discrete},
\begin{equation}
p_{i,j} \equiv \Tr[\ketbra{\beta_{i,j}}{\beta_{i,j}}\pmb{\rho}] = \sum_{\alpha \in \lambda_{i,j}} W_{\alpha}. 
\end{equation}
Here $p_{i,j}$ is the probability of projecting the state $\pmb{\rho}$ onto the bases vectors $\beta_{i,j}$ associated with the line $\lambda_{i,j}$. Different ways of making these mappings lead to different definitions of DWFs, i.e., a class of DWFs using the same fixed set of MUBs. Now, the resultant discrete Wigner functions at any phase space point $\alpha (q, p)$ can be found as \cite{gibbons2004discrete},
\begin{equation}
    \begin{aligned}
      W_{\alpha} = \frac{1}{d} Tr[ \textbf{A}_{\alpha} \pmb{\rho} ],
    \end{aligned}\label{DWFformula}
\end{equation}
where
\begin{equation}
    \begin{aligned}
      \textbf{A}_{\alpha} = \sum_{\lambda_{i,j} \ni \alpha} \textbf{P}_{i,j} - \textbf{I},
    \end{aligned}\label{A_formula}
\end{equation}
where $\textbf{P}_{i,j}$'s are the projectors associated with the lines $\lambda_{i,j}$'s. The operator $\textbf{A}_{\alpha}$ is known as the phase space point operator. 

The negative quantum states are obtained by considering the MUBs and striations to find the phase space point operators $\textbf{A}_{\alpha}$'s. The state corresponding to the normalized eigenvector of the minimum eigenvalue of $\textbf{A}_{\alpha}$ is known as the first negative quantum state, $\textit{i.e.}$, $NS_1$ state \cite{lalita2023harnessing}. Analogously, the second and third negative quantum states are represented by $NS_2$ state and $NS_3$  state, corresponding to the normalized eigenvectors of second and third negative eigenvalues of $\textbf{A}_{\alpha}$, respectively, and so on. A brief discussion of the striations, MUBs, and spectra of $\textbf{A}_{\alpha}$'s for two-qubit quantum systems is provided in Appendix \ref{appen-2-qubit}. The approximate explicit expressions of the two-qubit negative quantum states are given below,
\begin{eqnarray}
    \begin{aligned}
        \ket{NS_1} = \begin{pmatrix}
          -0.743\\ 
          -0.357(1 - i)\\
          0.102(1 + i)\\
          -0.414
        \end{pmatrix},\\
        \ket{NS_2} = \begin{pmatrix}
          0.789\\ 
          -0.289(1 - i)\\
          -0.289(1 + i)\\
          -0.211
        \end{pmatrix},\\
        \ket{NS_3} = \begin{pmatrix}
          -0.575\\ 
          -0.345(1 - i)\\
          -0.265(1 + i)\\
          0.575
        \end{pmatrix}.\\
    \end{aligned}
    \label{negative_quantum_states}
\end{eqnarray}
This work contemplates the quantum correlations, maximal fidelity, fidelity deviation, and success probability of two-qubit negative quantum states and the Bell state, particularly $\ket{\phi^{+}} = 1/\sqrt{2}(\ket{00} + \ket{11})$, under non-Markovian AD and RTN channels with(without) WM and QMR. 

\subsection{\label{noisy_channels} Noisy quantum channels}
When a quantum system $\pmb{\rho}$ interacts with the environment, its dynamical equation can be given as \cite{Kraus1971generalstatechange, Choi1975map}
\begin{equation}
\pmb{\rho}(t) = \sum_i \textbf{K}_{i}(t) \pmb{\rho}(0) \textbf{K}_{i}^{\dag}(t),
\end{equation}
where $\textbf{K}_i$'s are the Kraus operators, characterizing the noise and satisfying the completeness relation $\sum_{i} \textbf{K}_{i}^{\dag} \textbf{K}_{i} = 1$.
This type of representation is known as operator sum representation or the Kraus representation. We consider here the non-Markovian, unital, and non-unital channels. 
\subsubsection{Amplitude damping channel}
Amplitude-damping noise has been used to address a number of phenomena. Attenuation, energy dissipation, spontaneous photon emission, and idle errors in quantum computing in two-level systems are a few examples of these \cite{breuer2002theory, nielsen2010quantum, breuer2016colloquium}. The Kraus operators of the non-Markovian AD channel (non-unital) for a single qubit system are given as \cite{Bellomo2007NMAD},
\begin{equation}
    \mathbf{K_0^{AD}} = \begin{pmatrix}
     1 & 0\\
     0 & \sqrt{1 - \lambda(t)}
    \end{pmatrix},
    \mathbf{K_1^{AD}} = \begin{pmatrix}
    0 & \sqrt{\lambda(t)}\\
    0 & 0
\end{pmatrix},
\label{NMAD_Kraus_operators}
\end{equation}
where, $\lambda(t) = 1 - e^{-gt}\left(\frac{g}{l} \sinh{\frac{lt}{2}} + \cosh{\frac{lt}{2}}\right)^2$, and $l = \sqrt{g(g - 2\gamma)}$. The coupling strength $\gamma$ is related to the qubit relaxation time ($\tau_s = \frac{1}{\gamma}$), and $g$ is the line width that depends on the reservoir correlation time ($\tau_r = \frac{1}{g}$). The system exhibits Markovian and non-Markovian evolution of a state if $2\gamma << g$ and $2\gamma >> g $, respectively \cite{naikoo2019facets}. 

\subsubsection{\label{NMRTN_sec}Random Telegraph noise channel}
For a single qubit system, the Kraus operators of the non-Markovian RTN channel (unital) can be written as \cite{daffer2004depolarizing, nielsen2010quantum},
\begin{eqnarray}
      \mathbf{K_0^{RTN}} = \sqrt{\frac{1 + \Lambda(t)}{2}}\textbf{I},
      \mathbf{K_1^{RTN}} = \sqrt{\frac{1 - \Lambda(t)}{2}}\pmb{\sigma_z},
\label{NMRTN_Kraus_operators}      
\end{eqnarray}
where, $\Lambda(t)$ is the memory kernel given as,
\begin{equation}
      \Lambda(t) = e^{-\gamma t}\left[ \cos\left(\zeta \;\gamma t\right) + \frac{\sin\left(\zeta \;\gamma t\right)}{\zeta}\right].
\end{equation}
Here $\zeta = \sqrt{\left(\frac{2b}{\gamma}\right)^2 - 1}$, and $b$, $\gamma$ quantifies the RTN's system–environment coupling strength and fluctuation rate. Further, the $\textbf{I}$ and $\pmb{\sigma}_z$ are the identity and Pauli spin matrices, respectively. The dynamics is Markovian if {$(4 b \tau)^2 < 1$ and non-Markovian if $(4 b \tau)^2 > 1$ and here $\tau = \frac{1}{2\gamma}$ as discussed in \cite{kumar2018non}.

The evolution of a two-qubit system having local interactions under the non-Markovian AD and RTN noisy channels can be given as,
\begin{equation}
   \begin{aligned}
      \pmb{\rho}_{AB}(t) = \sum_{i = 0}^{1}\sum_{j = 0}^{1} (\mathbf{K}_i \otimes \mathbf{K}_j) \pmb{\rho}_{AB}(0) (\mathbf{K}_i \otimes \mathbf{K}_j)^{\dag},
      \end{aligned}
      \label{2qubitfinalrhot}
\end{equation}
the Kraus operators $\textbf{K}_0$, $\textbf{K}_1$ are the same as for a single qubit system under non-Markovian AD and RTN noise given by Eqs. (\ref{NMAD_Kraus_operators}) and (\ref{NMRTN_Kraus_operators}) respectively.

Now, we briefly outline a physical model for protecting quantum correlations and the universal quantum teleportation protocols of two-qubit negative quantum states using weak measurement and quantum measurement reversal. As shown in Fig. (\ref{schematic_With_WM_QMR}), Alice first performs weak measurement ($\pmb{M}_{\textit{WM}}(p_1, p_2)$), elaborated in Eq. (\ref{wm}), on the negative quantum states before distribution to Bob and Charlie via non-Markovian noisy quantum channels, discussed above. Bob and Charlie perform quantum measurement reversal ($\pmb{M}_{\textit{QMR}}(q_1, q_2)$), given in Eq. (\ref{qmr}), on receiving the qubits. The resulting state $\pmb{\rho}_f(t)$, Eq. (\ref{eq:WM_QMR}), can be made maximally entangled by choosing appropriate ($p_1, p_2$) and ($q_1, q_2$). Further, the state $\pmb{\rho}_f(t)$ can also be used for QT between Bob and Charlie.

\begin{figure}
    \centering
    \includegraphics[height=75mm,width=0.9\columnwidth]{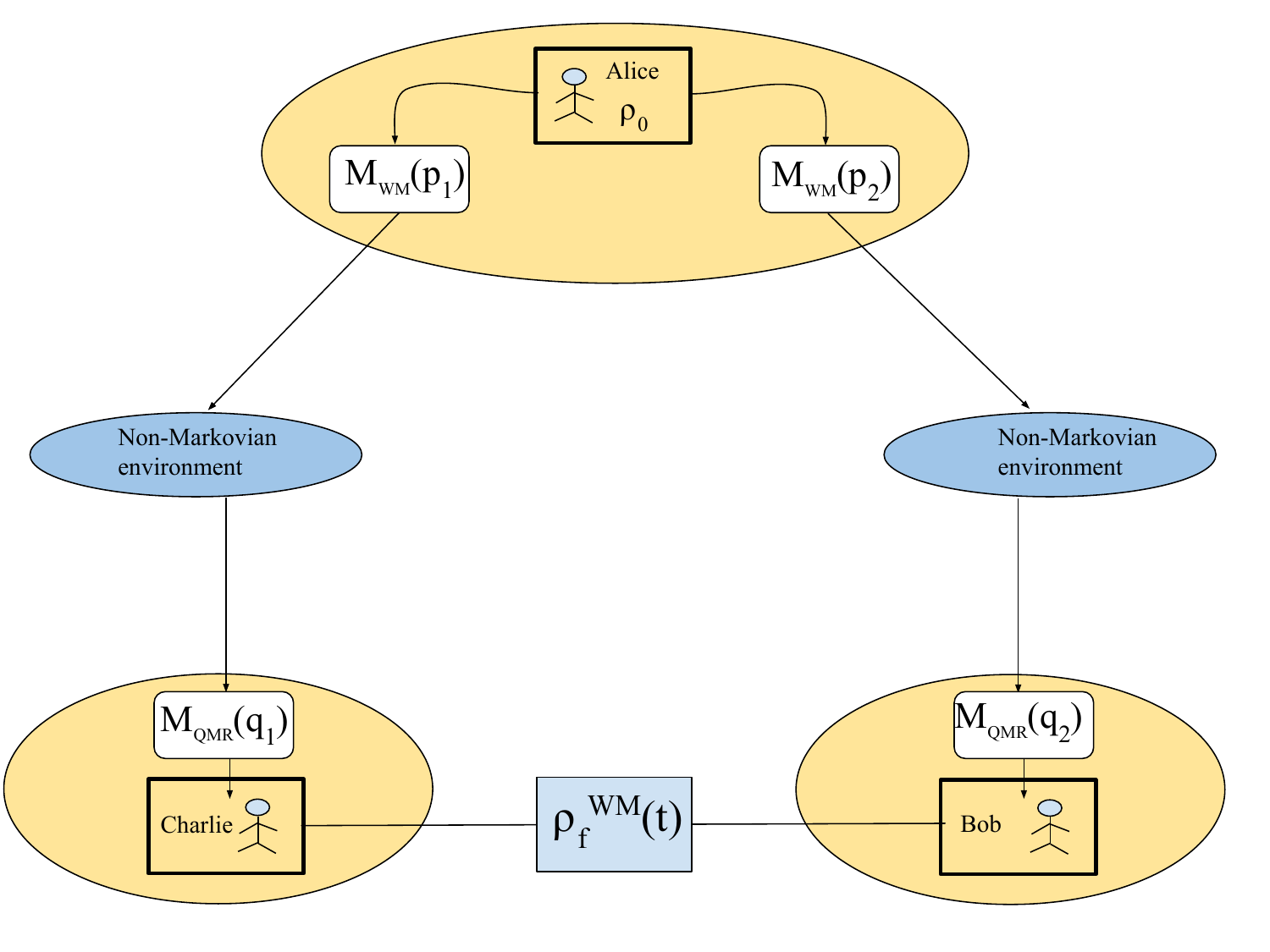}
    \caption{Schematic diagram for protecting quantum correlations of negative quantum states and Bell state using weak measurement ($\pmb{M}_{\textit{WM}}$) and quantum measurement reversal ($\pmb{M}_{\textit{QMR}}$).}
    \label{schematic_With_WM_QMR}
\end{figure}

The non-unitary WM and QMR operations are given as
\begin{equation}
   \begin{aligned}
     \pmb{M}_{\textit{WM}}(p_1, p_2) = \begin{pmatrix}
                                1 & 0\\
                                0 & \sqrt{1-p_1}
                               \end{pmatrix} \otimes \begin{pmatrix}
                                1 & 0\\
                                0 & \sqrt{1-p_2}
                               \end{pmatrix},
      \end{aligned}
      \label{wm}
\end{equation}
\begin{equation}
   \begin{aligned}
     \pmb{M}_{\textit{QMR}}(q_1, q_2) = \begin{pmatrix}
                                \sqrt{1-q_1} & 0\\
                                0 & 1
                               \end{pmatrix} \otimes \begin{pmatrix}
                                \sqrt{1-q_2} & 0\\
                                0 & 1
                               \end{pmatrix}.
      \end{aligned}
      \label{qmr}
\end{equation}

Here, $(p_1, p_2)$ and $(q_1, q_2)$ are the WM and QMR strength parameters, respectively. In our work, we have considered that ($p_1 = p_2 = p$) and ($q_1 = q_2 = q$), $\textit{i.e.}$, the strength of WM and QMR parameters are equal for both the qubits. It is important to note that via WM, the state does not collapse towards $\ket{00}$ or $\ket{11}$, indicating that appropriate operations, such as QMR, can still recover the measured state. After the sequential WM, non-Markovian channel, and QMR, the final state is
\begin{widetext}
\begin{equation}
 \pmb{\rho}_f(t) = \frac{\pmb{M}_{\textit{QMR}}\left( \sum_{i = 0}^{1}\sum_{j = 0}^{1} \mathbf{K}_{ij}[\pmb{M}_{\textit{WM}}\pmb{\rho}(0)\pmb{M}_{\textit{WM}}^{\dag}] \mathbf{K}_{ij}^{\dag}\right) \pmb{M}_{\textit{QMR}}^{\dag}}{P^{succ}},
 \label{eq:WM_QMR}
\end{equation}
\end{widetext}
here $\mathbf{K}_{ij} = (\mathbf{K}_i\otimes \mathbf{K}_j)$ are the Kraus operators of the non-Markovian noise. Since the WM and QMR are probabilistic in nature, $P^{succ} = Tr[\pmb{M}_{\textit{QMR}}\left( \sum_{i = 0}^{1}\sum_{j = 0}^{1} \mathbf{K}_{ij}[\pmb{M}_{\textit{WM}}\pmb{\rho}(0)\pmb{M}_{\textit{WM}}^{\dag}] \mathbf{K}_{ij}^{\dag}\right) \pmb{M}_{\textit{QMR}}^{\dag}]$ is their success probability \cite{pramanik2013improving, he2020enhancing}. A lower success probability of the protocol is the price for higher quantum correlations, maximal fidelity, and lesser fidelity deviations. 
\section{\label{protectingQCs}Protecting quantum correlations and universal quantum teleportation protocols under the non-Markovian AD and RTN channels using WM and QMR}
Here, we discuss the quantum correlations, maximal fidelity, fidelity deviation, and universal quantum teleportation protocols. We investigate the contribution of WM and QMR to the protection and enhancement of them. Further, we examine the variation of quantum concurrence, discord, steering, maximal fidelity, and fidelity deviation of the two-qubit negative quantum states, $\textit{i.e.}$, $NS_1, NS_2, NS_3$, and the Bell state under the non-Markovian AD and RTN channels, with(without) WM and QMR.

\subsection{Quantum correlations, maximal fidelity and fidelity deviation}
Quantum non-local correlations are one of the most remarkable and exclusive aspects of the quantum world with no equivalence in the classical world. Below, we discuss the quantum concurrence, discord, and steering in brief. Furthermore, maximal fidelity, fidelity deviation, and UQT protocols are discussed.
\subsubsection{Concurrence}
Entanglement is one of the most important sources of quantum information. Entanglement is also a fundamental component of quantum correlation in compound quantum systems.  For a two-qubit system, concurrence is an entanglement measure \cite{wootters1998entanglement}, which is defined as 
\begin{equation}
   \begin{aligned}
     C(\pmb{\rho}_{AB}) = \max \{0, \lambda_{1} - \lambda_{2} - \lambda_{3} - \lambda_{4}\},
      \end{aligned}
      \label{concur_eq.}
\end{equation}
Here $\lambda_{i}$'s are the eigenvalues of $\sqrt{\sqrt{\pmb{\rho}_{AB}} \tilde{\pmb{\rho}}_{AB} \sqrt{\pmb{\rho}_{AB}}}$, such that $\lambda_{1} \geq \lambda_{2} \geq \lambda_{3} \geq \lambda_{4}$, and $\Tilde{\pmb{\rho}}_{AB} = (\sigma_{y} \otimes \sigma_{y}) \pmb{\rho}_{AB}^{*} (\sigma_{y} \otimes \sigma_{y})$, where $\pmb{\rho}_{AB}^{*}$ is the complex conjugate of $\pmb{\rho}_{AB}$ and $\sigma_{y}$ is the Pauli bit-phase flip matrix.

\subsubsection{Discord}
There exist states that exhibit non-local behavior while remaining unentangled \cite{bennett1999quantum, luo2008quantum, ramkarthik2020quantum}. For quantifying such non-local correlations, Quantum Discord (QD) was developed  \cite{ollivier2001quantum, henderson2001classical}. It is a measure of a quantum system's overall non-local correlations. The QD for a bipartite composite quantum system $\pmb{\rho}_{AB}$, where $A$ and $B$ are the separate subsystems, is defined in the following manner,
\begin{equation}
    \begin{aligned}
        d(A:B) = t(A:B) - c(A|B),
    \end{aligned}
\end{equation}
where $t(A: B)$ and $c(A|B)$ are the total and classical correlations between the two subsystems, respectively. These correlations are defined as follows,

\begin{eqnarray}
    \nonumber
    t(A:B) &=& s(A) + s(B) - s(A, B),\\
    c(A|B) &=& s(A) - s(A|B).
    \label{corr_von-neumann_eqn}
\end{eqnarray}

Here $s(A) = -Tr(\pmb{\rho}_A ln \pmb{\rho}_A )$ and $s(B) = -Tr(\pmb{\rho}_B ln \pmb{\rho}_B )$ represent the von Neumann entropies of subsystem states $\pmb{\rho}_A$ and $\pmb{\rho}_B$ respectively. The $s(A, B)$ and $s(A|B)$ are the system's mutual and conditional quantum entropies, respectively, and are given by,
\begin{eqnarray}
    \nonumber
    s(A, B) &=& -Tr(\pmb{\rho}_{AB} ln \pmb{\rho}_{AB}),\\
    s(A|B) &=&  min_{\{\pmb{\pi}_i\}}\sum_{i=1}^{\mathbf{H}_B} p_i s(\pmb{\rho}_{A|\pmb{\pi}_i}).
    \label{conditional_entropy}
\end{eqnarray}

Here, $\mathbf{H}_B$ is the Hilbert space dimension of subsystem B, and minimization is performed over all possible measurement operators $\pmb{\pi}_i$. The post-measurement state for subsystem $A$ when a measurement is performed on subsystem $B$ is $\pmb{\rho}_{A|\pmb{\pi}_i}$ and $p_i = Tr(\pmb{\pi}_i^{\dag} \pmb{\pi}_i \pmb{\rho}_{AB})$ is the probability of measurement operators $\pmb{\pi}_i$. It is possible to write the state $\pmb{\rho}_{A|\pmb{\pi}_i}$ explicitly as,
\begin{equation}
    \begin{aligned}
    \pmb{\rho}_{A|\pmb{\pi}_i} = \frac{1}{p_i} Tr(\pmb{\pi}_i \pmb{\rho}_{AB}\pmb{\pi}_i)
    \end{aligned}
\end{equation}

It is important to keep in mind that we can determine the QD by measuring either subsystem $A$ or $B$. In this case, we are measuring subsystem $B$ and restricting the measurement to one qubit because the depreciation in Eq. (\ref{conditional_entropy}) depends on $2^m$ parameters of the measurement operators $i$, where $m$ is the number of qubits in subsystem $B$. The general measurement parameters for $m = 1$ are 
\begin{eqnarray}
    \nonumber
    \pmb{\pi}_1 &=& I_A \otimes \ket{l}_{BB} \bra{l},\\
    \pmb{\pi}_2 &=& I_A \otimes \ket{m}_{BB} \bra{m},
\end{eqnarray}
where $\ket{l} = \cos{\frac{\theta}{2}}\ket{0} + e^{i\phi}\sin{\frac{\theta}{2}}\ket{1}$ and $\ket{m} = \sin{\frac{\theta}{2}}\ket{0} - e^{i\phi}\sin{\frac{\theta}{2}}\ket{1}$, and $0 \leq \theta \leq \pi$, $0 \leq \phi \leq 2\pi$. 
Using Eq. (\ref{corr_von-neumann_eqn}), the discord can be defined in terms of von Neumann, joint von Neumann, and conditional quantum entropies as

\begin{equation}
    \begin{aligned}
        d(A:B) = s(B) - s(A, B) + s(A|B).
    \end{aligned}
    \label{discord_von_neumann_entropy_eqn}
\end{equation}

\subsubsection{\label{steering}Steering}
The concept of steering was established in  \cite{schrodinger1935discussion, schrodinger1936probability}. If Alice and Bob share an entangled pair, Alice can remotely steer Bob's state by performing measurements exclusively on her half of the system. This type of quantum correlation is referred to as steering or EPR steering. It lies between the Bell non-locality \cite{brunner2014bell} and entanglement \cite{horodecki2009quantum}. It is also a resource of quantum teleportation \cite{fan2022quantum}. By considering how much a steering inequality is maximally violated, we can determine the degree of steerability of a particular quantum state \cite{costa2016quantification}. For two-qubit systems, the steering formula is given as,
\begin{equation}
   \begin{aligned}
     S_{n}(\pmb{\rho}_{AB}) = \max \left\{0, \frac{\Omega_{n} - 1}{\sqrt{n} - 1}\right\}.
    \end{aligned}
    \label{steering_eq.}
\end{equation}
When $n = 2, 3$ per party measurements are involved, called two (three)-measurement steering, respectively $\Omega_2 = \sqrt{c^2 - c_{min}^2}$, and $\Omega_3 = c$. Here $c = \sqrt{\textbf{c}^2}$, and $c_{min} \equiv min\{|c_i|\}$, $c_i$'s are the eigenvalues of correlation matrix $\textbf{T} = \{t_{ij}\}$, and $t_{ij} = Tr[\pmb{\rho}_{AB} (\pmb{\sigma_{i}} \otimes \pmb{\sigma_{j}})]$.

\subsubsection{\label{maxiaml fidelity and fidelity deviation sec.}Maximal fidelity and fidelity deviation}
A fundamental protocol for transmitting quantum information using shared entanglement and local operations and classical communication (LOCC) is quantum teleportation \cite{bennett1993teleporting}. The maximal average fidelity $(F_{\pmb{\rho}_{AB}})$ \cite{horodecki1996teleportation, badziag2000local}, and deviation in the fidelity $(\Delta_{\pmb{\rho}_{AB}})$ \cite{bang2018fidelity, ghosal2020optimal} are usually used to determine the quality of a teleportation protocol. 

The maximal average fidelity (or maximal fidelity) is the maximal of all average fidelity obtained by strategies using the standard protocol and local unitary operations. For two-qubit states with $det(\textbf{T}) < 0$ (here $\textbf{T}$ is the correlation matrix discussed in Sec. (\ref{steering}), the maximal average fidelity can be determined as \cite{horodecki1996teleportation, ghosal2020optimal}
\begin{equation}
   \begin{aligned}
     F_{\pmb{\rho}_{AB}} = \frac{1}{2} \left( 1 + \frac{1}{3} \sum_{i = 1}^{3}|e_{i}|\right),
    \end{aligned}
    \label{max_fid_eq.}
\end{equation}
here $e_{i}$'s are the eigenvalues of the correlation matrix $\textbf{T}$.

Fidelity deviation is described as the standard deviation of fidelity values across all possible input states. For a two-qubit state with $det(\textbf{T}) < 0$, fidelity deviation corresponding to the optimal protocol is \cite{bang2018fidelity, ghosal2020optimal} 
\begin{equation}
   \begin{aligned}
     \Delta_{\pmb{\rho}_{AB}} = \frac{1}{3\sqrt{10}} \sqrt{\sum_{i < j =1}^{3} (|e_{i}| - |e_{j}|)^2},
    \end{aligned}
    \label{FD_eq.}
\end{equation}
here also $e_{i}$'s are the eigenvalues of the correlation matrix $\textbf{T}$.

When $F_{\pmb{\rho}_{AB}} > \frac{2}{3}$, the particular two-qubit state $\pmb{\rho}_{AB}$ is beneficial for quantum teleportation (QT) \cite{horodecki1996teleportation, horodecki1999general}; here $\frac{2}{3}$ is the highest average fidelity possible with classical protocols. However, $\pmb{\rho}_{AB}$ is universal for quantum teleportation (UQT) iff $\Delta_{\pmb{\rho}_{AB}} = 0$ and $F_{\pmb{\rho}_{AB}} > \frac{2}{3}$. From these relations, some conditions are formulated in \cite{ghosal2020optimal} that can be checked for a two-qubit state $\pmb{\rho}_{AB}$ to verify its validity for universal quantum teleportation. These conditions are:
\\$(i)$ The states with the property $det(\textbf{T}) < 0$ form a subset of the states that are beneficial for QT; here, $\textbf{T}$ is the correlation matrix. 
\\$(ii)$ A two-qubit state $\pmb{\rho}_{AB}$ is useful for UQT iff, $|e_1| = |e_2| = |e_3| > 1/3$; here $|e_1|, |e_2|, |e_3|$ are the eigenvalues of the correlation matrix $\textbf{T}$.

All input states will be transported with the same fidelity if the two-qubit state satisfies the aforementioned universality condition. Just like the Bell state, all the considered two-qubit negative quantum states also have $det(\textbf{T}) < 0$. Subsequently, all the considered two-qubit negative quantum states approximately satisfy $|e_1| = |e_2| = |e_3| > 1/3$ except the $NS_2$ state. Whereas, the $NS_2$ state have $|e_1| = |e_2| \neq |e_3| > 1/3$.

We will now examine the performance of two-qubit negative quantum states for quantum correlations and UQT protocols with(without) WM and QMR under non-Markovian AD and RTN channels. The WM and QMR strength parameters $p$ and $q$ are optimized at $t = 0$ to obtain maximum concurrence for the two-qubit $NS_1$, $NS_2$, $NS_3$, and the Bell states. In the presence of non-Markovian AD and RTN channels, the optimal combinations for the $NS_1$, $NS_2$, $NS_3$, and the Bell states are ($p = 0.17$, $q = 0.54$), ($p = 0.05$, $q = 0.74$), ($p = 0.05$, $q = 0.05$), and ($p = 0.01$, $q = 0.01$), respectively. Furthermore, we use the same set of optimal parameters to study the quantum discord, steering, and UQT protocols. Also, these parameters are fixed during the evolution of the states under the non-Markovian AD and RTN channels.

\subsection{Concurrence under non-Markovian AD and RTN channels, with(without) WM and QMR}
We study the evolution of concurrence of two-qubit negative quantum states and the Bell state under the non-Markovian AD and RTN channels. Further, the effect of WM and QMR on the concurrence of two-qubit negative quantum states and the Bell state under the non-Markovian AD and RTN channel is analyzed.

\subsubsection{Under non-Markovian AD noise}
Figure (\ref{concur_NMAD}a) illustrates the behavior of the concurrence of considered two-qubit negative quantum states and the Bell state without WM and QMR under the non-Markovian AD noise. It can be seen that the $NS_3$ state has entanglement equal to the Bell state, and both states have the highest entanglement in the initial period. But for a longer duration, the $NS_3$ state dominates the Bell state and other two-qubit negative quantum states under the non-Markovian AD noise.

Upon optimizing WM and QMR strength parameters, the variation of concurrence is depicted in Fig. (\ref{concur_NMAD}b). The WM and QMR can be seen to significantly improve and protect the concurrence of the $NS_1$ and $NS_2$ states. In fact, with WM and QMR, the $NS_2$ state shows entanglement equal to the Bell state and the $NS_3$ state at $t = 0$. Moreover, over time, the $NS_2$ state shows concurrence more than all considered states. Additionally, the concurrence of the Bell state and $NS_3$ state remain unaltered with WM and QMR under non-Markovian AD evolution. It is also clear from Fig. (\ref{concur_NMAD}b) that the two-qubit negative quantum states dominate the Bell state in terms of preserving their entanglement under non-Markovian AD noise for a longer duration.

\begin{figure}
    \centering
    \includegraphics[height=45mm,width=1\columnwidth]{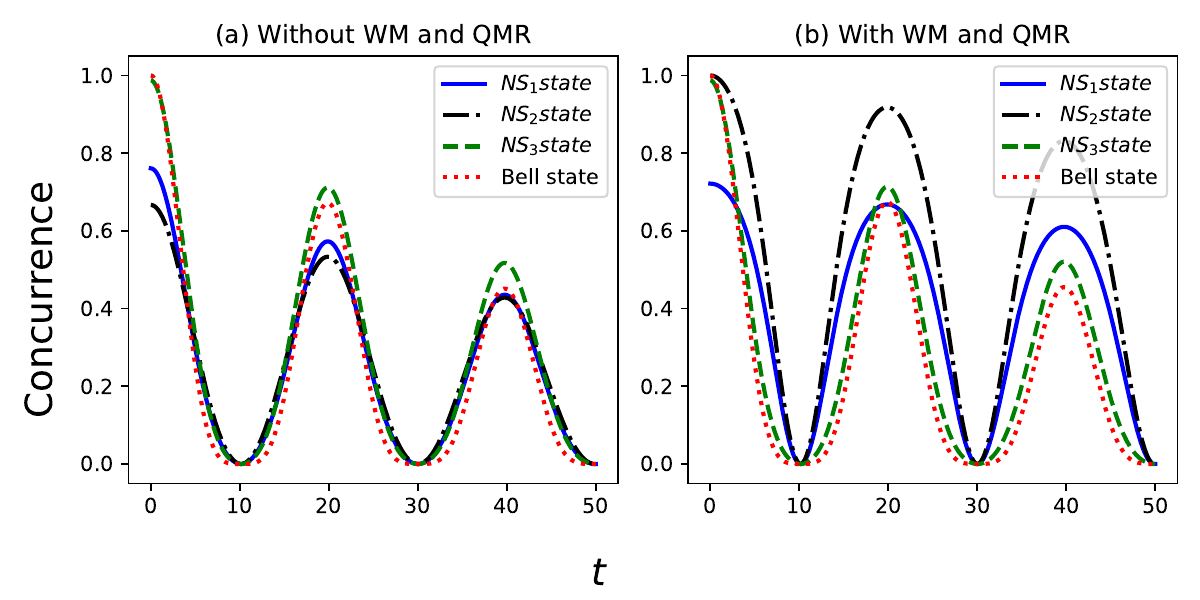}
    \caption{Variation of concurrence of $NS_1$, $NS_2$, $NS_3$, and Bell state under non-Markovian AD channel without WM and QMR in subplot (a), and with WM and QMR in subplot (b) with time. Here, for $NS_1$ ($p = 0.17$, $q = 0.54$), for $NS_2$ ($p = 0.05$, $q = 0.74$), for $NS_3$ ($p = 0.05$, $q = 0.05$), and for Bell state ($p = 0.01$, $q = 0.01$). The non-Markovian AD channel parameters are $g = 0.01$ and $\gamma = 5$.}
    \label{concur_NMAD}
\end{figure}

\subsubsection{Under non-Markovian RTN channel}
The evolution of the concurrence in the absence of WM and QMR under the non-Markovian RTN channel is shown in Fig. (\ref{concur_NMRTN}a). Under the non-Markovian RTN channel, the $NS_3$ state shows entanglement variations almost equal to the Bell state, and both have the highest entanglement over time. All the states show expectedly oscillatory, slowly decaying behavior under the non-Markovian RTN channel.

The concurrence variation in the presence of WM and QMR under the non-Markovian RTN channel can be seen in Fig. ($\ref{concur_NMRTN}b$). The concurrence of the Bell state and $NS_3$ state remain unaltered with WM and QMR under non-Markovian RTN evolution. The concurrence of the $NS_1$ and $NS_2$ states have been improved. In fact, the $NS_2$ state exhibits entanglement comparable to that of the Bell state and $NS_3$ state over time when WM and QMR are employed.
\begin{figure}[!htpb]
    \centering
    \includegraphics[height=45mm,width=1\columnwidth]{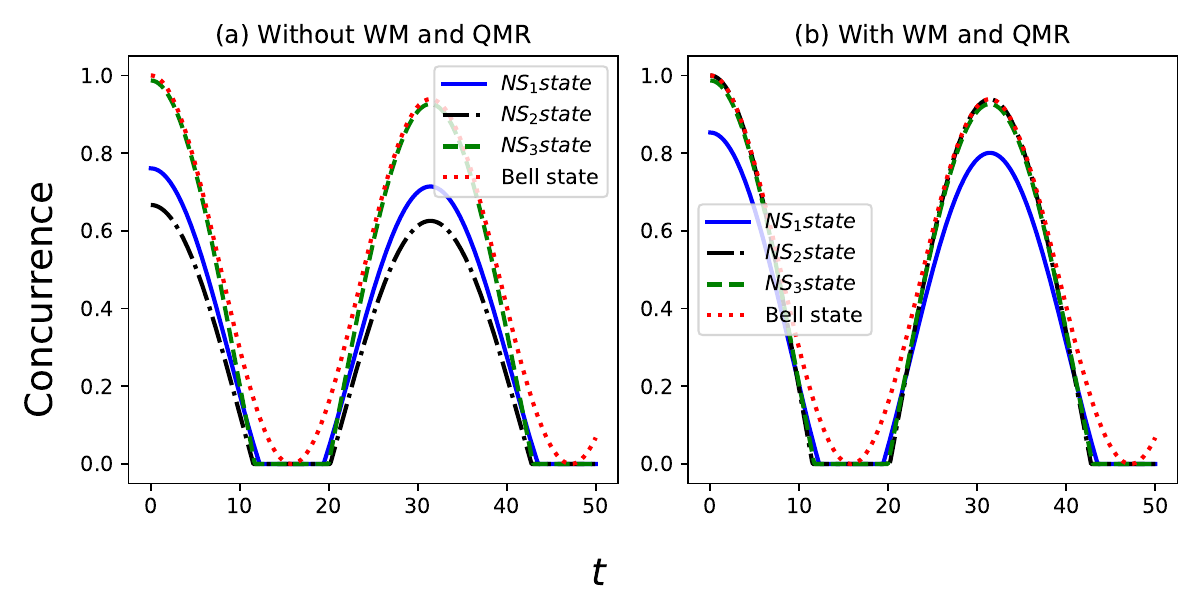}
    \caption{Variation of concurrence of $NS_1$, $NS_2$, $NS_3$, and Bell state under non-Markovian RTN channel without WM and QMR in subplot (a), and with WM and QMR in subplot (b) with time. Here, for $NS_1$ ($p = 0.17$, $q = 0.54$), for $NS_2$ ($p = 0.05$, $q = 0.74$), for $NS_3$ ($p = 0.05$, $q = 0.05$), and for Bell state ($p = 0.01$, $q = 0.01$). The non-Markovian RTN channel parameters are $b = 0.05$ and $\gamma = 0.001$.}
    \label{concur_NMRTN}
\end{figure}

\subsection{Discord under non-Markovian AD and RTN channels, with(without) WM and QMR}
To study the dynamics of discord under non-Markovian AD and RTN channels, Eq. (\ref{discord_von_neumann_entropy_eqn}) and Eqs. (\ref{2qubitfinalrhot}, \ref{NMAD_Kraus_operators}, \ref{NMRTN_Kraus_operators}) are employed. Further, Eq. (\ref{eq:WM_QMR}) is utilized to understand the effect of WM and QMR on discord dynamics. 

\subsubsection{Under non-Markovian AD channel}
Figure (\ref{discord_NMAD}a) shows the variation of discord of two-qubit $NS_1$, $NS_2$, $NS_3$, and the Bell state under non-Markovian AD noise without WM and QMR. The variation of discord of $NS_3$ state is similar to the Bell state under non-Markovian AD noise. Also, these states have the highest discord values among all the considered states.

The variation of discord under the non-Markovian AD noise with WM and QMR of the two-qubit $NS_1$, $NS_2$, $NS_3$, and the Bell state is depicted in Fig. (\ref{discord_NMAD}b). The discord of the $NS_1$ and $NS_2$ states can be seen to be enhanced by the WM and QMR. The $NS_2$ state exhibits discord at $t = 0$ comparable to the Bell state and the $NS_3$ state with WM and QMR. In addition, the $NS_2$ state shows more discord over time than all other considered states. Furthermore, during non-Markovian AD evolution, the Bell state discord with WM and QMR remains unaffected. 

\begin{figure}[!htpb]
    \centering
    \includegraphics[height=45mm,width=1\columnwidth]{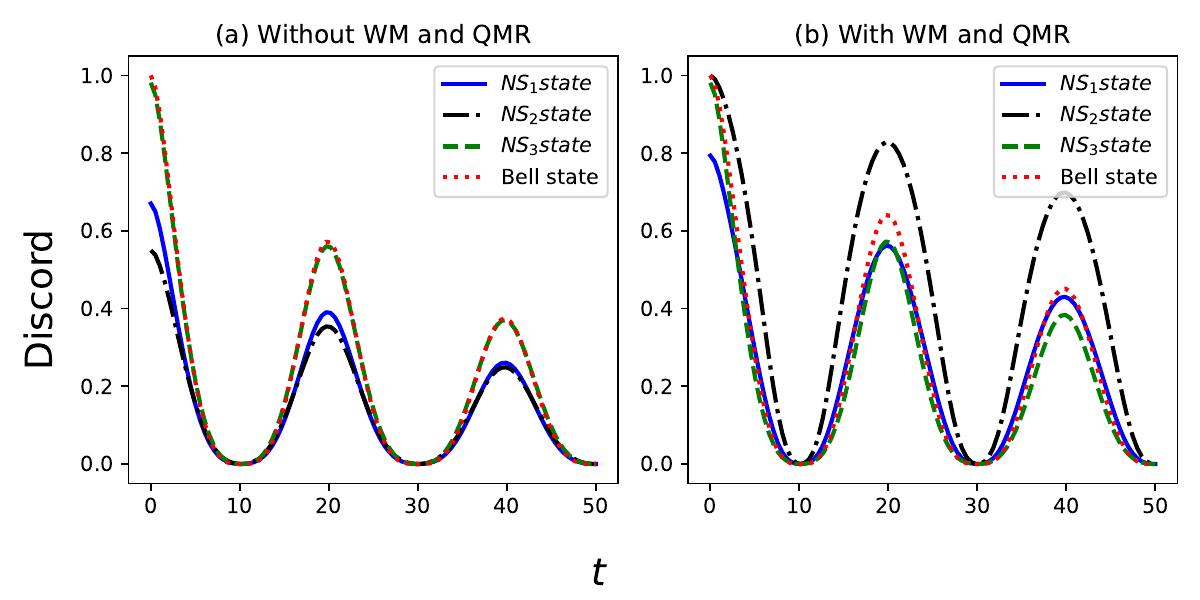}
    \caption{Variation of discord of $NS_1$, $NS_2$, $NS_3$, and Bell state under non-Markovian AD channel without WM and QMR in subplot (a), and with WM and QMR in subplot (b) with time. Here, for $NS_1$ ($p = 0.17$, $q = 0.54$), for $NS_2$ ($p = 0.05$, $q = 0.74$), for $NS_3$ ($p = 0.05$, $q = 0.05$), and for Bell state ($p = 0.01$, $q = 0.01$). The non-Markovian AD channel parameters are $g = 0.01$, and $\gamma = 5$.}
    \label{discord_NMAD}
\end{figure}

\subsubsection{Under non-Markovian RTN channel}
Figure (\ref{discord_NMRTN}a) depicts that without WM and QMR, at $t = 0$, the $NS_3$ state's discord is equal to the Bell state. It dominates the Bell state and other considered states over time under the non-Markovian RTN channel.

When the two-qubit $NS_1$, $NS_2$, $NS_3$, and Bell state are subjected to the non-Markovian RTN channel and WM and QMR, the variations in discord are represented by Fig. (\ref{discord_NMRTN}b). At $t = 0$, the discord of $NS_1$ and $NS_2$ state is improved to a reasonable extent. In fact, with WM and QMR, at $t = 0$, the $NS_2$ state shows discord equal to the Bell state and $NS_3$ state. It also dominates all other considered states over time. At the same time, the Bell state and $NS_3$ state remain uninfluenced by WM and QMR.

\begin{figure}[!htpb]
    \centering
    \includegraphics[height=45mm,width=1\columnwidth]{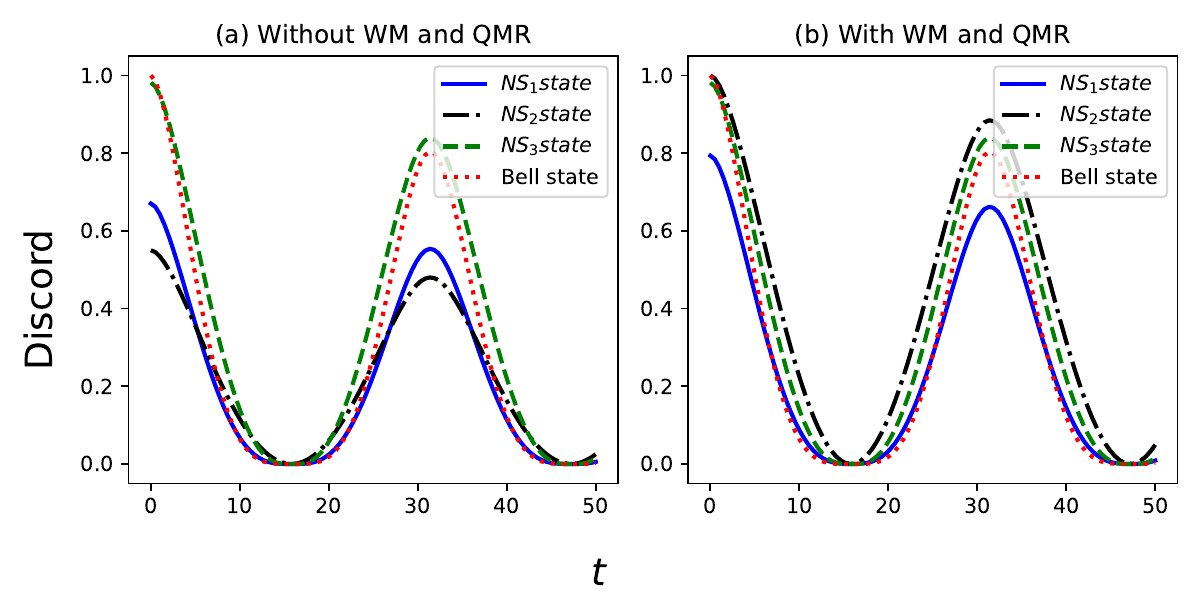}
    \caption{Variation of discord of $NS_1$, $NS_2$, $NS_3$, and Bell state under non-Markovian RTN channel without WM and QMR in subplot (a), and with WM and QMR in subplot (b) with time. Here, for $NS_1$ ($p = 0.17$, $q = 0.54$), for $NS_2$ ($p = 0.05$, $q = 0.74$), for $NS_3$ ($p = 0.05$, $q = 0.05$), and for Bell state ($p = 0.01$, $q = 0.01$). The non-Markovian RTN channel parameters are $b = 0.05$ and $\gamma = 0.001$.}
    \label{discord_NMRTN}
\end{figure}

\subsection{Steering under non-Markovian AD and RTN channels, with(without) WM and QMR}
To investigate the dynamics of two (three)-measurement steering of two-qubit $NS_1$, $NS_2$, $NS_3$, and the Bell states under non-Markovian AD and RTN channel, Eqs. (\ref{2qubitfinalrhot}, \ref{NMAD_Kraus_operators}, \ref{NMRTN_Kraus_operators}) and (\ref{steering_eq.}) for $n =2$, $3$ are utilized.  To comprehend how the WM and QMR affect the two (three)-measurement steering of the Bell state and the negative quantum states of the two-qubit system under non-Markovian AD and RTN noise, the Eq. ({\ref{eq:WM_QMR}}) is utilized. 

\subsubsection{Under non-Markovian AD channel}
At $t = 0$, the three-measurement steering of the $NS_1$ and $NS_2$ states is higher than their two-measurement steering, whereas their values are the same for the $NS_3$ state and the Bell state as shown in Figs. (\ref{steering_NMAD}a) and (\ref{steering_NMAD}c). The $NS_3$ state and the Bell state show a maximum of two (three)-measurement steering for the initial period, whereas the Bell state dominates for a longer duration without WM and QMR. 

The two (three)-measurement steering of the $NS_1$ and $NS_2$ states can be seen to have improved significantly with WM and QMR, from Figs. (\ref{steering_NMAD}b) and (\ref{steering_NMAD}d). In fact, with WM and QMR, the $NS_2$ state shows two (three)-measurement steering even higher than the Bell state over time. Furthermore, the two (three)-measurement steering of the Bell state remains unaffected. 

\begin{figure}[!htpb]
    \centering
    \includegraphics[height=80mm,width=1\columnwidth]{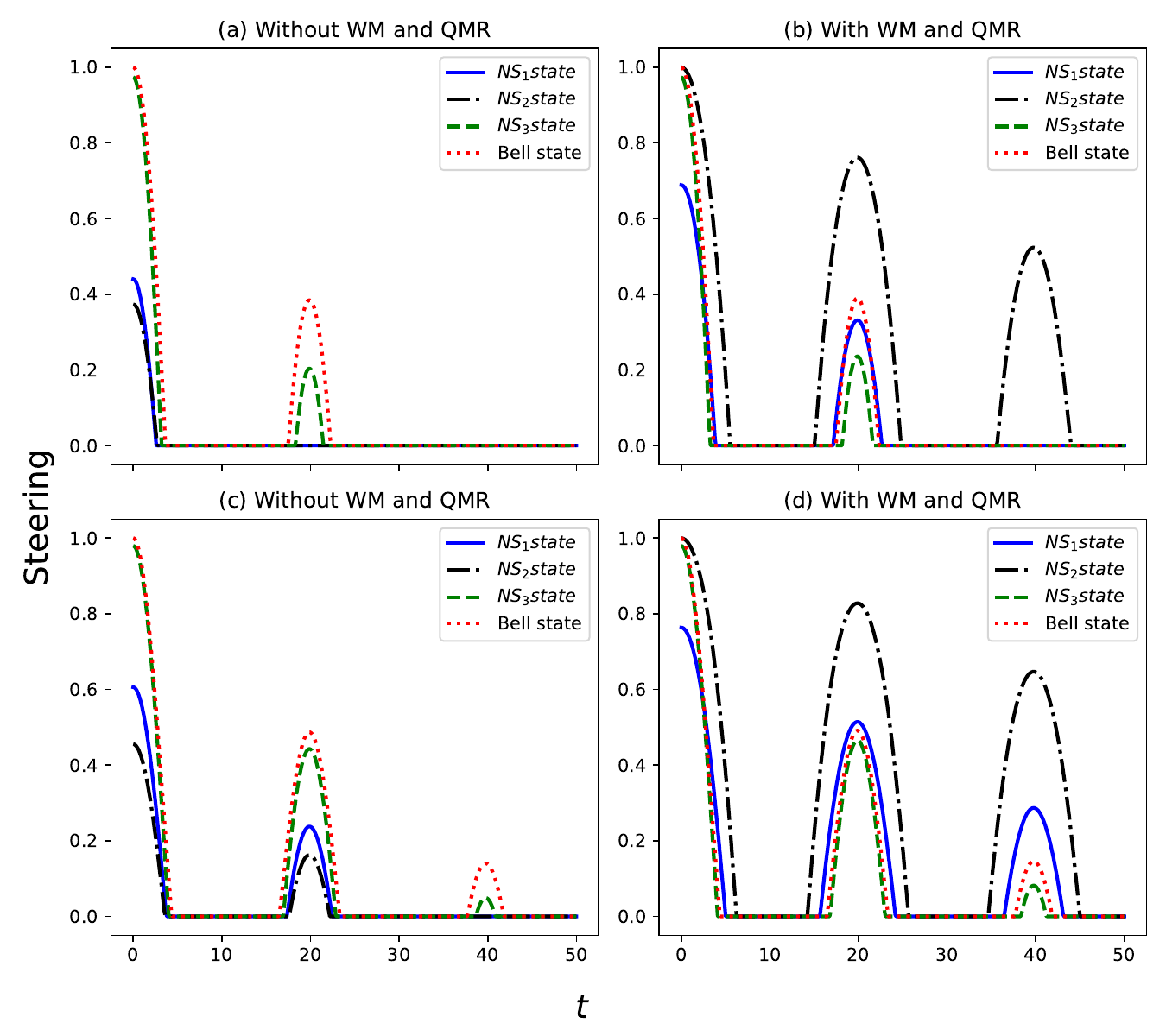}
    \caption{Variation of two (three)-measurement steering of $NS_1$, $NS_2$, $NS_3$, and Bell state under non-Markovian AD channel with time. Here, subplots (a) and (c) represent the two-measurement and three-measurement steering without WM and QMR, and subplots (b) and (d) represent the two-measurement and three-measurement steering with WM and QMR, respectively. Here, for $NS_1$ ($p = 0.17$, $q = 0.54$), for $NS_2$ ($p = 0.05$, $q = 0.74$), for $NS_3$ ($p = 0.05$, $q = 0.05$), and for Bell state ($p = 0.01$, $q = 0.01$). The non-Markovian AD channel parameters are $g = 0.01$, and $\gamma = 5$.}
    \label{steering_NMAD}
\end{figure}

\subsubsection{Under non-Markovian RTN channel}
The decay of two (three)-measurement steering of the two-qubit negative quantum states and the Bell state under the non-Markovian RTN channel is lesser than the non-Markovian AD case. Like the non-Markovian AD case, the $NS_3$ state and the Bell state show a maximum of two (three)-measurement steering values with time under the non-Markovian RTN channel, which can be seen from Figs. (\ref{steering_NMRTN}a) and (\ref{steering_NMRTN}b) respectively. The three-measurement steering of $NS_1$ and $NS_2$ state at $t = 0$ is higher than their two-measurement steering.

The variations in two (three)-measurement steering of the $NS_1$, $NS_2$, $NS_3$, and the Bell state when subjected to non-Markovian RTN channel with WM and QMR are shown in Figs. (\ref{steering_NMRTN}c) and (\ref{steering_NMRTN}d). From Figs. (\ref{steering_NMRTN}c) and (\ref{steering_NMRTN}d), it is evident that the two (three)-measurement steering of the $NS_1$ and $NS_2$ states have improved using WM and QMR. The $NS_2$ state now exhibits two (three)-measurement steering equivalent to the $NS_3$ state with WM and QMR over time. Moreover, there is no change in the two (three)-measurement steering of the $NS_3$ and Bell states.

\begin{figure}[!htpb]
    \centering
    \includegraphics[height=80mm,width=1\columnwidth]{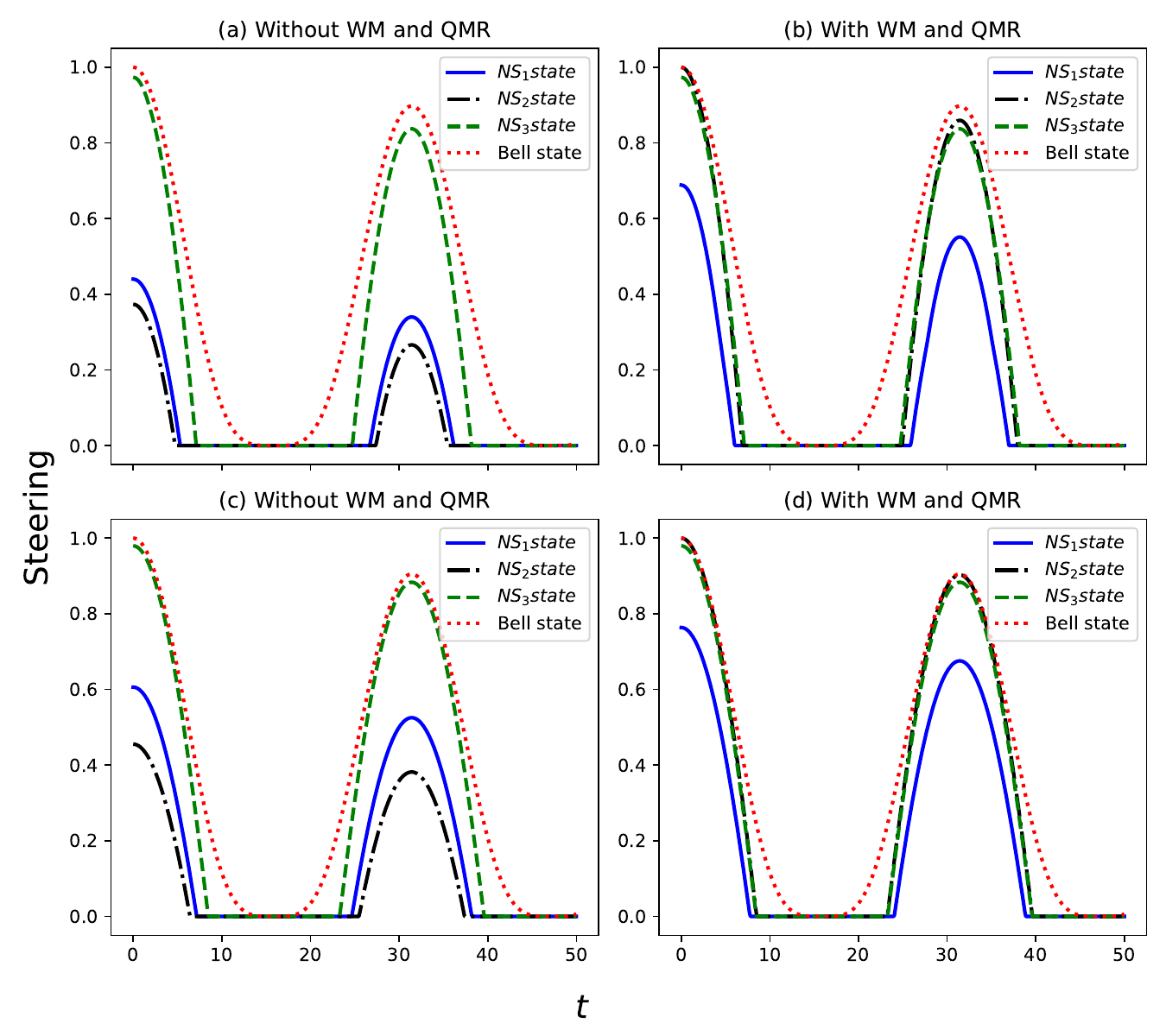}
    \caption{Variation of two (three)-measurement steering of $NS_1$, $NS_2$, $NS_3$, and Bell state under non-Markovian RTN channel with time. Here, subplots (a) and (b) represent the two-measurement and three-measurement steering without WM and QMR, and subplots (b) and (d) represent the two-measurement and three-measurement steering with WM and QMR, respectively. Here, for $NS_1$ ($p = 0.17$, $q = 0.54$), for $NS_2$ ($p = 0.05$, $q = 0.74$), for $NS_3$ ($p = 0.05$, $q = 0.05$), and for Bell state ($p = 0.01$, $q = 0.01$). The non-Markovian RTN channel parameters are $b = 0.05$ and $\gamma = 0.001$.}
    \label{steering_NMRTN}
\end{figure}

\subsection{Maximal fidelity under non-Markovian AD and RTN channels, with(without) WM and QMR}
We explore the dynamics of maximal fidelity of the two-qubit $NS_1$, $NS_2$, and $NS_3$, and the Bell state under non-Markovian AD and RTN channel. Further, Eq. (\ref{eq:WM_QMR}) is employed to understand the impact of the WM and QMR on the maximal fidelity in the presence of the non-Markovian AD and RTN channels.

\subsubsection{Under non-Markovian AD channel}
From Fig. (\ref{Maxiaml Fidelity_NMAD}a), it is clear that at $t = 0$, the Bell state and $NS_3$ state attain a maximal fidelity value of $1$. But, the Bell state's maximal fidelity sustains longer with time. However, just like the Bell state, all the considered two-qubit negative quantum states have maximal fidelity always greater than $\frac{2}{3}$ under non-Markovian AD noise.

The WM and QMR successfully enhance and protect the maximal fidelity of the two-qubit $NS_1$ and $NS_2$ states. Indeed, the $NS_1$ and $NS_2$ state's fidelity leads the Bell state, as depicted in Fig. (\ref{Maxiaml Fidelity_NMAD}b). On the other hand, the fidelity variations of the two-qubit $NS_3$, and the Bell state remain unaltered with WM and QMR.
\begin{figure}[!htpb]
    \centering
    \includegraphics[height=45mm,width=1\columnwidth]{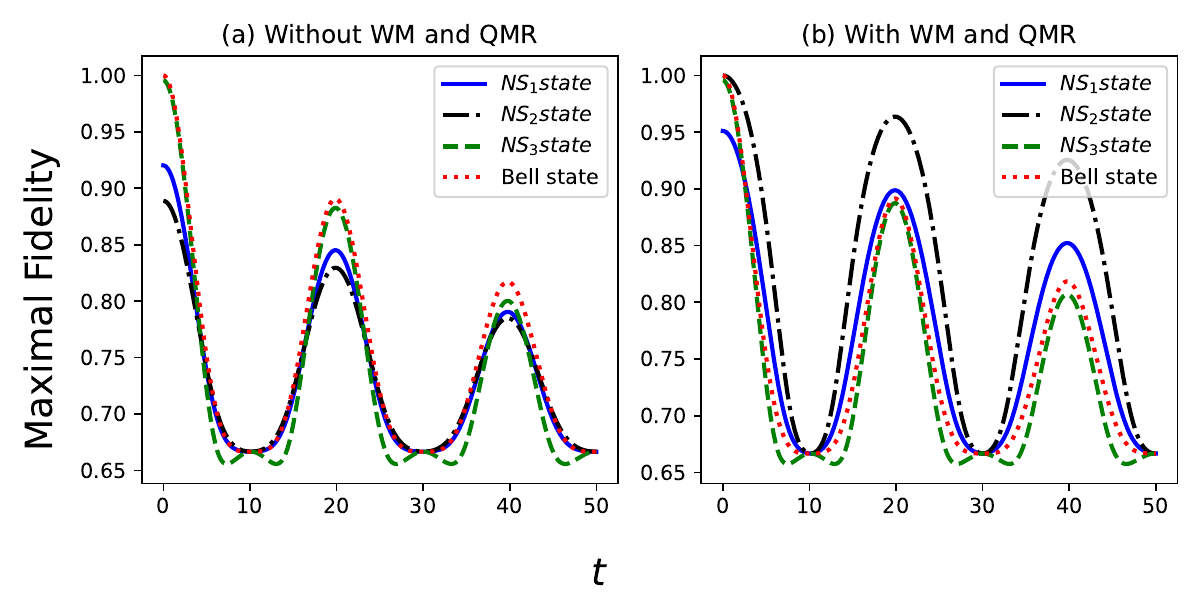}
    \caption{Variation of maximal average fidelity of $NS_1$, $NS_2$, $NS_3$, and Bell state under non-Markovian AD channel without WM and QMR in subplot (a), and with WM and QMR in subplot (b) with time. Here, for $NS_1$ ($p = 0.17$, $q = 0.54$), for $NS_2$ ($p = 0.05$, $q = 0.74$), for $NS_3$ ($p = 0.05$, $q = 0.05$), and for Bell state ($p = 0.01$, $q = 0.01$). The non-Markovian AD channel parameters are $g = 0.01$ and $\gamma = 5$.}
    \label{Maxiaml Fidelity_NMAD}
\end{figure}

\subsubsection{Under non-Markovian RTN channel}
The two-qubit $NS_2$ and Bell states show similar maximal fidelity behavior under the non-Markovian RTN channel for short evolution times. The decay in maximal fidelity over time of all the considered states is gradual compared to the non-Markovian AD noise, as shown in Fig. (\ref{Maximal Fidelity_NMRTN}a).

As depicted in Fig. (\ref{Maximal Fidelity_NMRTN}b), the maximal fidelity variations of the $NS_3$, and the Bell states, when subjected to non-Markovian RTN channel, remain unaffected by WM and QMR. On the other hand, there is a significant improvement in the maximal fidelity of $NS_1$ and $NS_2$ states with WM and QMR. In fact, with WM and QMR, the $NS_2$ state's maximal fidelity variations are comparable to the $NS_3$ state.  

\begin{figure}[!htpb]
    \centering
    \includegraphics[height=45mm,width=1\columnwidth]{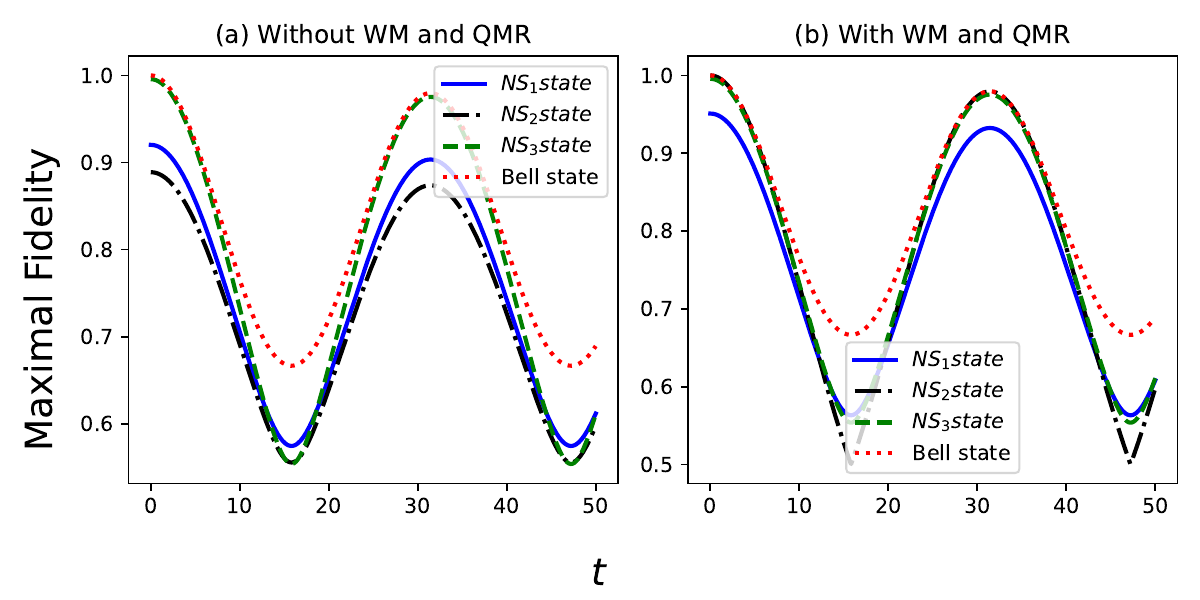}
    \caption{Variation of maximal average fidelity of $NS_1$, $NS_2$, $NS_3$, and Bell state under non-Markovian RTN channel without WM and QMR in subplot (a), and with WM and QMR in subplot (b) with time. Here, for $NS_1$ ($p = 0.17$, $q = 0.54$), for $NS_2$ ($p = 0.05$, $q = 0.74$), for $NS_3$ ($p = 0.05$, $q = 0.05$), and for Bell state ($p = 0.01$, $q = 0.01$). The non-Markovian RTN channel parameters are $b = 0.05$ and $\gamma = 0.001$.}
    \label{Maximal Fidelity_NMRTN}
\end{figure}

\subsection{Fidelity deviation under non-Markovian AD and RTN channels, with(without) WM and QMR}
We analyze variation in the fidelity deviation of the Bell state and two-qubit negative quantum states in the presence of non-Markovian AD and RTN noise. Moreover, we study the impact of the WM and QMR on the fidelity deviation of the above-mentioned states.

\subsubsection{Under non-Markovian AD channel}
Figure (\ref{FD_NMAD}a) represents the variations in fidelity deviation of the states mentioned above under non-Markovian AD noise. We can observe from Fig. (\ref{FD_NMAD}a) that at $t = 0$, all the considered states except the $NS_2$ state show fidelity deviation approximately equal to zero. The reason for this is discussed in Sec. \ref{maxiaml fidelity and fidelity deviation sec.} that, except the $NS_2$ state all the negative quantum states have $|e_1| = |e_2| = |e_3| > 1/3$. Moreover, all the two-qubit negative quantum states and the Bell state show oscillatory variations in fidelity deviation in synchronization, except for some kinks between the oscillations where the negative quantum states are approaching minimum deviation in fidelity. In these regions, the behavior of the Bell state is different from the two-qubit negative quantum states. If we compare the variation in fidelity deviation of all the states under the non-Markovian AD noise, the $NS_3$ state exhibits a smaller fidelity deviation than the other considered states. This shows that without WM and QMR, the $NS_3$ state is relatively better than the Bell state and other negative quantum states for UQT under non-Markovian AD noise.

The WM and QMR are seen to successfully reduce the deviation in the fidelity of all the considered two-qubit negative quantum states. The WM and QMR are able to squeeze the non-zero fidelity deviation area of two-qubit $NS_1$ and $NS_2$ states in contrast to the $NS_3$ and Bell states, as depicted in Fig. (\ref{FD_NMAD}b). This makes them more suitable candidates for UQT under non-Markovian AD noise.

\begin{figure}[!htpb]
    \centering
    \includegraphics[height=45mm,width=1\columnwidth]{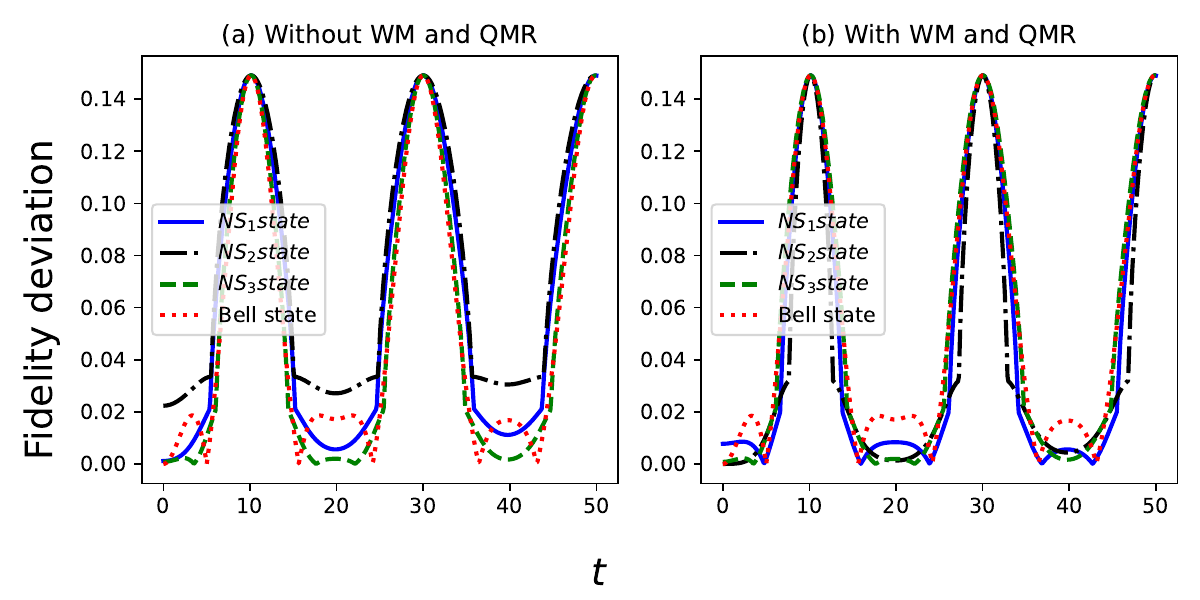}
    \caption{Variation of fidelity deviation of $NS_1$, $NS_2$, $NS_3$, and Bell state under non-Markovian AD channel without WM and QMR in subplot (a), and with WM and QMR in subplot (b) with time. Here, for $NS_1$ ($p = 0.17$, $q = 0.54$), for $NS_2$ ($p = 0.05$, $q = 0.74$), for $NS_3$ ($p = 0.05$, $q = 0.05$), and for Bell state ($p = 0.01$, $q = 0.01$). The non-Markovian AD channel parameters are $g = 0.01$ and $\gamma = 5$.}
    \label{FD_NMAD}
\end{figure}

\subsubsection{Under non-Markovian RTN channel}
Under the non-Markovian RTN noise, all the two-qubit negative quantum states show less fidelity deviation than the Bell state, as shown in Fig. (\ref{FD_NMRTN}a). Thus, compared to the Bell state, they are more suitable for UQT.

With the WM and QMR, the $NS_2$ state shows zero deviation in fidelity under the non-Markovian RTN channel, making it an ideal state for UQT as demonstrated in Fig. (\ref{FD_NMRTN}b). On the other hand, the WM and QMR have no impact on the fidelity deviation of the Bell state and $NS_3$ state under the non-Markovian RTN. Additionally, the deviation in the fidelity of $NS_1$ state is reduced to some extent, as depicted by Fig. (\ref{FD_NMRTN}b).

\begin{figure}[!htpb]
    \centering
    \includegraphics[height=45mm,width=1\columnwidth]{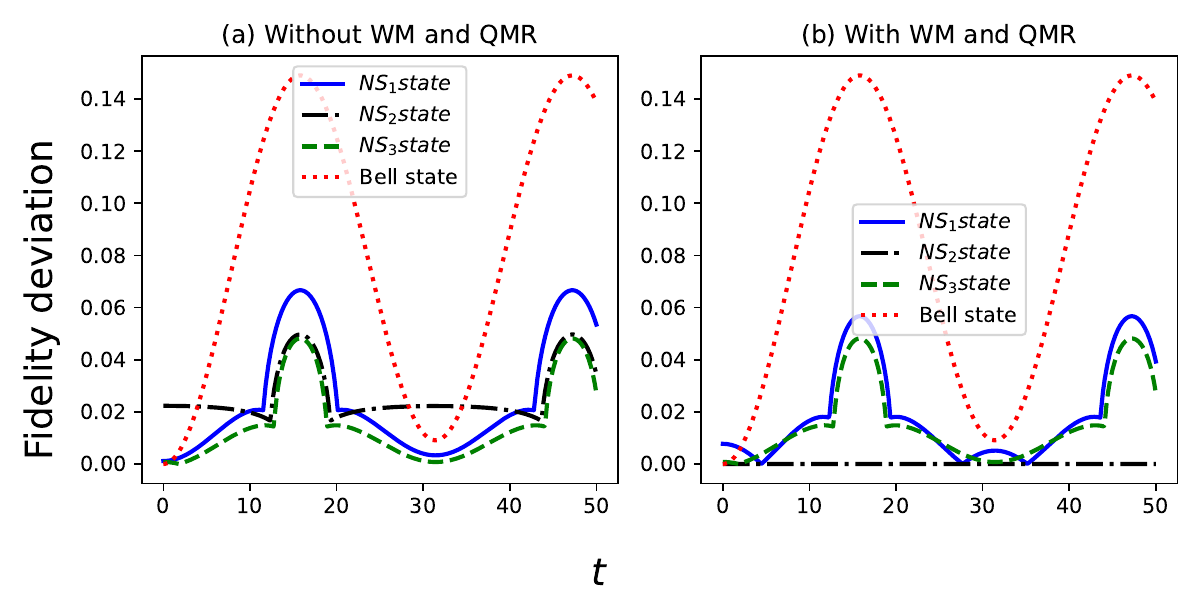}
    \caption{Variation of fidelity deviation of $NS_1$, $NS_2$, $NS_3$, and Bell state under non-Markovian RTN channel without WM and QMR in subplot (a), and with WM and QMR in subplot (b) with time. Here, for $NS_1$ ($p = 0.17$, $q = 0.54$), for $NS_2$ ($p = 0.05$, $q = 0.74$), for $NS_3$ ($p = 0.05$, $q = 0.05$), and for Bell state ($p = 0.01$, $q = 0.01$). The non-Markovian RTN channel parameters are $b = 0.05$ and $\gamma = 0.001$.}
    \label{FD_NMRTN}
\end{figure}

\begin{table}
\begin{tabular}{ | m{2.5cm}| m{3cm} | m{3cm} |}
  \hline 
  \textbf{Quantum correlations and UQT requirements} & \textbf{Without WM and QMR} &  \textbf{With WM and QMR}\\ 
  \hline
  Concurrence & \scriptsize$NS_3>BS>NS_1>NS_2$ & \scriptsize$NS_2>NS_1>NS_3>BS$ \\
  \hline
  Discord & \scriptsize$NS_3=BS>NS_1>NS_2$ & \scriptsize$NS_2>NS_1 \approx BS>NS_3$\\
  \hline
  Two (three)-measurement steering & \scriptsize$BS>NS_3>NS_1>NS_2$ & \scriptsize$NS_2>NS_1>BS>NS_3$\\
  \hline
  Maximal Fidelity & \scriptsize$BS>NS_3>NS_1>NS_2$ & \scriptsize$NS_2>NS_1>BS>NS_3$\\
  \hline
  Fidelity deviation & \scriptsize$NS_3<BS<NS_1<NS_2$ & \scriptsize$NS_2 \approx NS_1<NS_3<BS$\\
  \hline
\end{tabular}
\caption{\label{table1} Comparison of the quantum correlations, maximal fidelity, and fidelity deviation variations of two-qubit $NS_1$, $NS_2$, $NS_3$, and the Bell state (BS) under the non-Markovian AD channel ($t>0$).}
\end{table}

\begin{table}
\begin{tabular}{ | m{2.5cm}| m{3cm} | m{3cm} |}
  \hline 
  \textbf{Quantum correlations and UQT requirements} & \textbf{Without WM and QMR} &  \textbf{With WM and QMR}\\ 
  \hline
  Concurrence & \scriptsize$NS_3 \approx BS>NS_1>NS_2$ & \scriptsize$NS_2 \approx NS_3 \approx BS>NS_1$ \\
  \hline
  Discord & \scriptsize$NS_3>BS>NS_1>NS_2$ & \scriptsize$NS_2>NS_3>BS>NS_1$\\
  \hline
  Two (three)-measurement steering & \scriptsize$BS>NS_3>NS_1>NS_2$ & \scriptsize$BS>NS_2 \approx NS_3>NS_1$\\
  \hline
  Maximal Fidelity & \scriptsize$BS>NS_3>NS_1>NS_2$ & \scriptsize$BS>NS_2 \approx NS_3> NS_1$\\
  \hline
  Fidelity deviation & \scriptsize$NS_3<NS_1<NS_2<BS$ & \scriptsize$NS_2<NS_3<NS_1<BS$\\
  \hline
\end{tabular}
\caption{\label{table2} Comparison of the quantum correlations, maximal fidelity, and fidelity deviation variations of two-qubit $NS_1$, $NS_2$, $NS_3$, and the Bell state (BS) under the non-Markovian RTN channel ($t>0$).}
\end{table}

\section{\label{result&discussion}Results and discussion}
The WM and QMR strength parameters are optimized at time $t = 0$ for two-qubit $NS_1$, $NS_2$, $NS_3$, and the Bell states and optimal combinations of $(p, q)$ are obtained. If we keep the same values of $(p, q)$ for all the states, without picking up the optimal ones for every state, their behavior is depicted in Fig. (\ref{concur_NMAD_same_p_and_q}). It can be seen that the Bell state need not provide the best results. Moreover, by doing this, the main advantage of having WM and QMR would be lost. This is so because these parameters are in the control of the experimentalist and can be leveraged to get the optimal quantum correlations and UQT requirements for the states under consideration. This motivates the need to optimize the WM and QMR strength parameters at time $t = 0$ for two-qubit $NS_1$, $NS_2$, $NS_3$, and the Bell states. Under non-Markovian noisy quantum channels, the effect of WM and QMR on the quantum correlations and maximal fidelity of the two-qubit states is greater when the WM and QMR strength parameters are large, as discussed in \cite{sun2017recovering}, at the cost of success probability. This is corroborated by Fig. (\ref{P_success}). The WM and QMR also minimize the fidelity deviation of non-maximally entangled two-qubit states, consistent with  \cite{sabale2023towards}. We discuss these criteria below in conjunction with our results. The trade-off relation between success probability, discussed in Sec. (\ref{Model}), and WM and QMR strength parameters ($p, q$) can be observed in Fig. (\ref{P_success}) under non-Markovian AD and RTN channels, respectively. 

Due to WM and QMR, the two-qubit $NS_2$ state's (non-maximally entangled state at time $t = 0$) concurrence, discord, steering, maximal fidelity, and fidelity deviation are seen to improve significantly as we can observe that the strength parameters ($p$, $q$) are highest for this particular state with some non-zero finite success probability. The $NS_2$ state's concurrence, discord, steering, and maximal fidelity were observed to be higher than all other considered states under non-Markovian AD noise. This pattern was followed by the behavior of discord under non-Markovian RTN noise. Also, there is a notable improvement in the concurrence, steering, and maximal fidelity of the $NS_2$ state under the non-Markovian RTN channel which is attributed to the WM and QMR. In fact, under the non-Markovian RTN channel, the $NS_2$ state's concurrence, steering, and maximal fidelity were seen to be equivalent to the $NS_3$ state. The WM and QMR reduced the $NS_2$ state's fidelity deviation under non-Markovian AD and RTN channels. In fact, under the non-Markovian RTN channel, its fidelity deviation is almost zero with time, making it an ideal candidate for UQT.

The WM and QMR positively impact the two-qubit $NS_1$ state's quantum correlations and UQT requirements under non-Markovian AD and RTN channels. Moreover, with WM and QMR, this state maintains its concurrence, steering, maximal fidelity, and fidelity deviation to higher values than the $NS_3$ and Bell states under non-Markovian AD noise at the expense of low success probability. Whereas under the non-Markovian RTN channel, its fidelity deviation is less than the Bell state with(without) WM and QMR. Moreover, the fidelity deviation of all the considered two-qubit negative quantum states is less than the Bell state under the non-Markovian RTN channel, with(without) WM and QMR. 

The WM and QMR strength parameters are quite low for two-qubit $NS_3$ and Bell states with high success probability, which makes sense as the two-qubit $NS_3$ and Bell states are maximally entangled states at time $t = 0$. This explains why the quantum correlations, maximal fidelity, and fidelity deviation of the $NS_3$ and Bell states were not changed with WM and QMR. Additionally, the fidelity deviation of $NS_3$ state is lesser than the Bell state with(without) WM and QMR, making it a more suitable candidate for UQT.

The overall pattern hierarchy of quantum correlations, maximal fidelity, and fidelity deviation with(without) WM and QMR under non-Markovian AD and RTN channels is summarized in TABLE \ref{table1} and TABLE \ref{table2} respectively, for optimal combinations of WM and QMR strength parameters as follows $NS_1$ ($p = 0.17$, $q = 0.54$), $NS_2$ ($p = 0.05$, $q = 0.74$), $NS_3$ ($p = 0.05$, $q = 0.05$), and Bell state ($p = 0.01$, $q = 0.01$) under both non-Markovian AD and RTN channels.

\begin{figure}
    \centering
    \includegraphics[height=75mm,width=1\columnwidth]{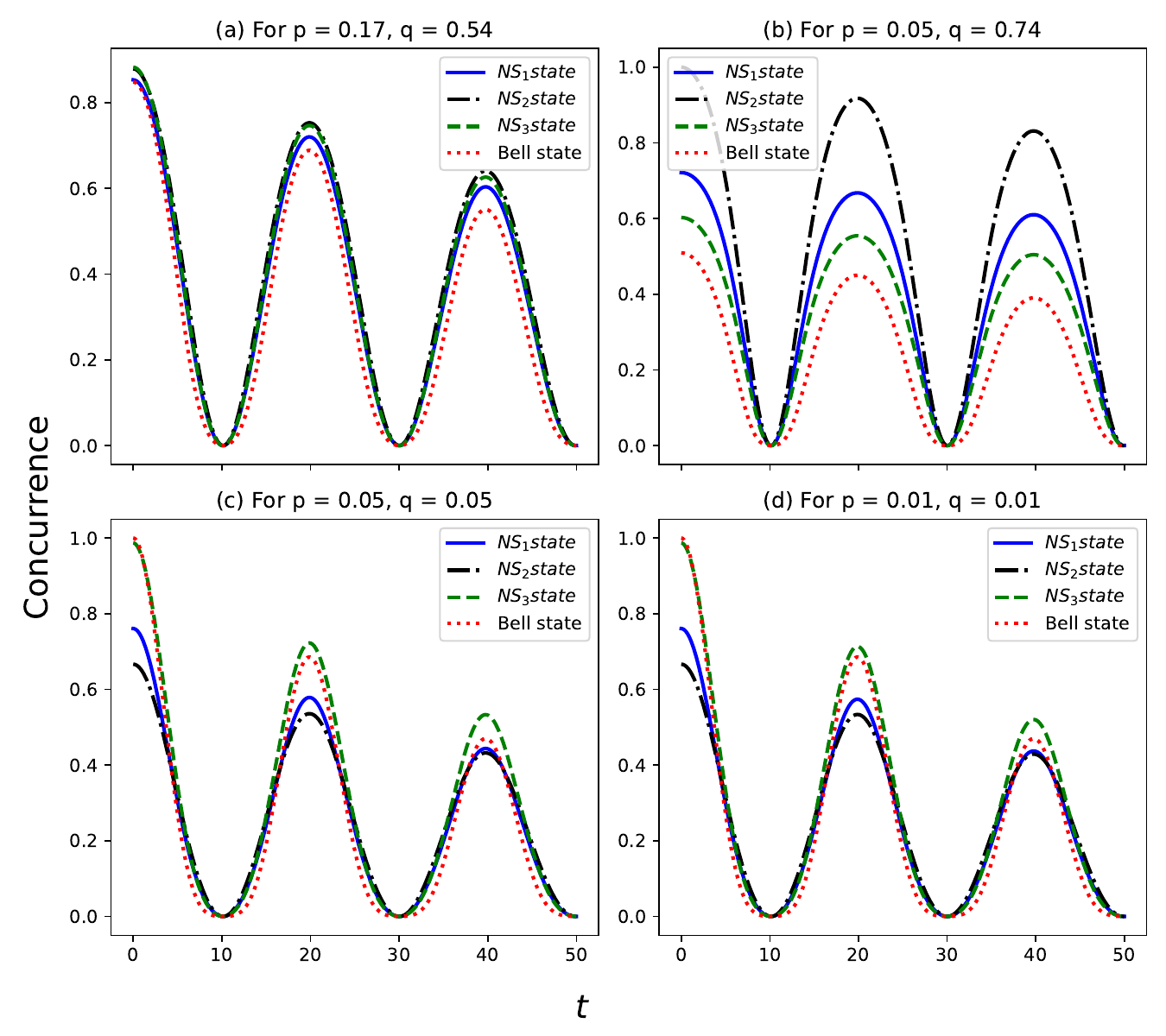}
    \caption{Variation of concurrence of $NS_1$, $NS_2$, $NS_3$, and Bell state under non-Markovian AD channel with WM and QMR in subplots (a), (b), (c), and (d). In subplot (a) $p = 0.17$, $q = 0.54$ for all states, in subplot (b) $p = 0.05$, $q = 0.74$ for all states, in subplot (c) $p = 0.05$, $q = 0.05$ for all states, and in subplot (d) $p = 0.01$, $q = 0.01$ for all states. The non-Markovian AD channel parameters are $g = 0.01$ and $\gamma = 5$.}
    \label{concur_NMAD_same_p_and_q}
\end{figure}

\begin{figure}[!htpb]
    \centering
    \includegraphics[height=50mm,width=1\columnwidth]{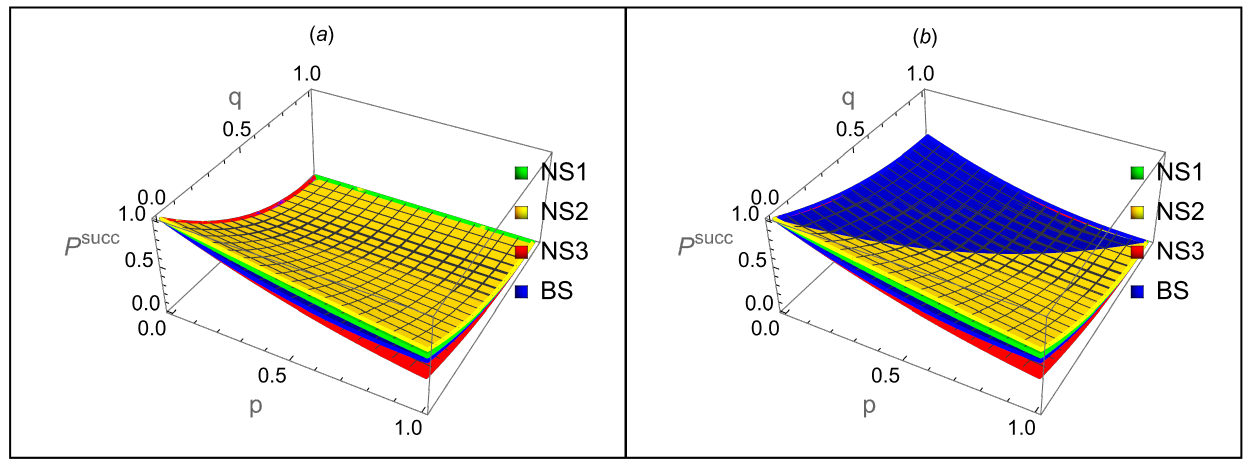}
    \caption{Variation of success probability of $NS_1$, $NS_2$, $NS_3$, and Bell state ($BS$) under non-Markovian AD and RTN channels in subplots (a) and (b), respectively. Here, the figures are depicted for WM strength ($p$) and QMR strength ($q$) at time $t = 10$. The non-Markovian AD and RTN channel parameters are ($g = 0.01$, $\gamma = 5$) and ($b = 0.05$, $\gamma = 0.001$), respectively.}
    \label{P_success}
\end{figure}

\section{\label{conclusion}Conclusion}
In this article, we have investigated the impact of weak measurement (WM) and quantum measurement reversal (QMR) on the quantum correlations and universal quantum teleportation (UQT) of two-qubit $NS_1$, $NS_2$, $NS_3$, and the Bell states under both non-unital (non-Markovian Amplitude Damping) and unital (non-Markovian Random Telegraph Noise) quantum channels. To this end, we discussed the negative quantum states followed by their quantum correlations, particularly concurrence, discord, and steering, and their maximal fidelity, and fidelity deviation under the influence of noisy quantum channels with(without) WM and QMR. It was shown that WM and QMR brought out better performance of the negative quantum states for the quantum correlations and the UQT. Additionally examined is the relationship of trade-offs between weak measurement and quantum measurement reversal parameters and success likelihood. Interestingly, it was observed that some surpassed the Bell state's performance over time. It was also observed that the fidelity deviation of negative quantum states was lesser compared to the Bell state with(without) WM and QMR when states evolved through unital quantum channels. It was found that the $NS_3$ state has a lower fidelity deviation than the Bell state with(without) WM and QMR with high success probability, making it a better choice for UQT.

However, with WM and QMR, the two-qubit $NS_2$ state showed zero fidelity deviation under the unital channel. Under the non-unital channel, this state fares better among all the considered states, making it an ideal state for UQT. Also, the $NS_1$ state, with the weak measurement and its reversal, outperforms the Bell state under the non-unital channel for UQT. These properties of the two-qubit negative quantum states affected by WM and QMR can be useful in protocols involving quantum correlations and UQT.

\section*{Acknowledgements}
SB acknowledges the support from the Interdisciplinary Cyber-Physical Systems (ICPS programme of the Department of Science and Technology (DST), India, Grant No.: DST/ICPS/QuST/Theme-1/2019/6).
\appendix
\section{\label{appen-2-qubit}Two-qubit discrete Wigner functions}
A $4 \times 4$ array defines the discrete phase space for two-qubit quantum systems. The Galois field elements $\mathcal{F}_4 = \{0, 1, \omega, \omega^2\}$ labels the points in this discrete phase space. This phase space has five potential sets of parallel lines (striations) shown in Fig. (\ref{striation3}).
\begin{figure}[!htpb]
    \centering
    \includegraphics[width = 0.4\textwidth, height = 65mm]{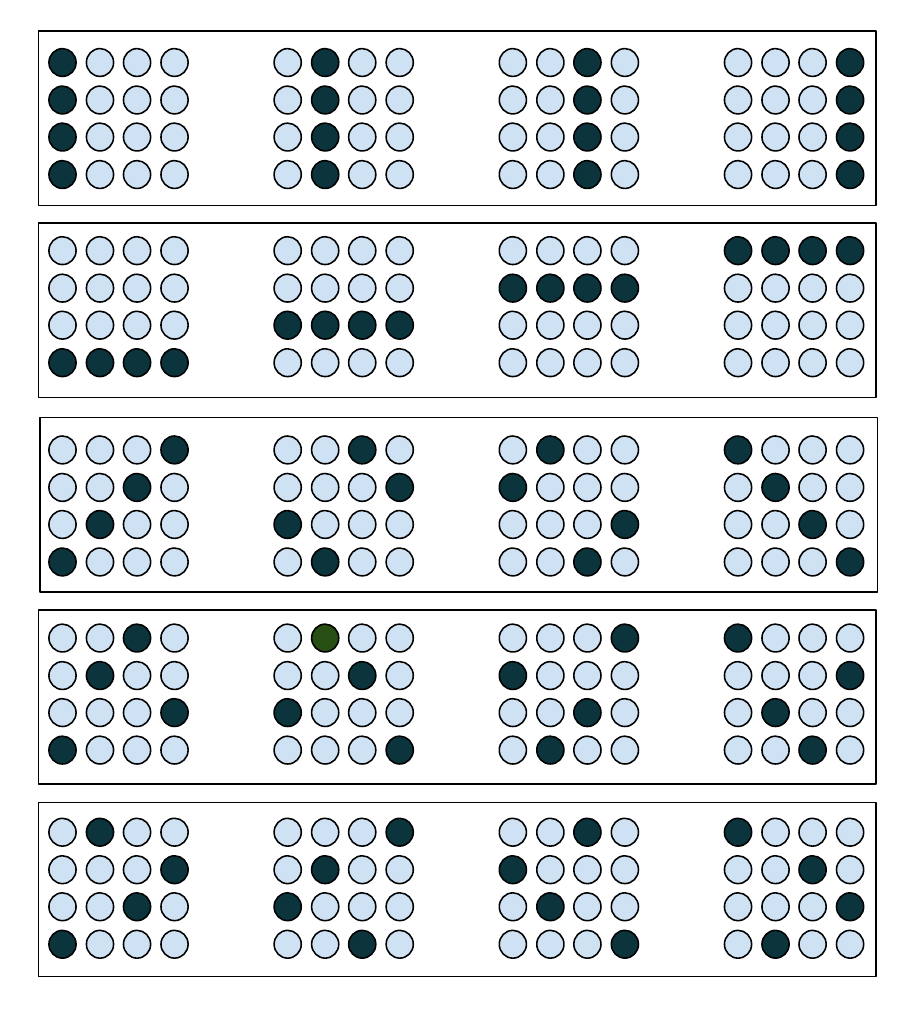}
    \caption{Lines and striations of the $4 \times 4$ phase space.}
    \label{striation3}
\end{figure}
A set of five MUBs is required for a one-to-one mapping with striations (given in TABLE \ref{table3}).  
\begin{table}
\begin{center}
\begin{tabular}{ | m{2cm}| m{6cm} | }

  \hline
  \textbf{Striation} & \textbf{MUBs associated with striation}\\ 
  \hline
  1 &  
  $\begin{pmatrix}
  1\\ 
  0\\
  0\\
  0
\end{pmatrix}$,
$\begin{pmatrix}
  0\\ 
  1\\
  0\\
  0
\end{pmatrix}$,
$\begin{pmatrix}
  0\\ 
  0\\
  1\\
  0
\end{pmatrix}$,
$\begin{pmatrix}
  0\\ 
  0\\
  0\\
  1
\end{pmatrix}$\\
\hline
  2 &  
  $\frac{1}{2}\begin{pmatrix}
  1\\ 
  1\\
  1\\
  1
\end{pmatrix}$,
$\frac{1}{2}\begin{pmatrix}
  1\\ 
 -1\\
  1\\
  -1
\end{pmatrix}$,
$\frac{1}{2}\begin{pmatrix}
  1\\ 
  1\\
  -1\\
  -1
\end{pmatrix}$,
$\frac{1}{2}\begin{pmatrix}
  1\\ 
  -1\\
  -1\\
  1
\end{pmatrix}$\\
\hline
  3 &  
$\frac{1}{2}\begin{pmatrix}
  1\\ 
  -i\\
   i\\
   1
\end{pmatrix}$,
$\frac{1}{2}\begin{pmatrix}
  1\\ 
  i\\
   i\\
   -1
\end{pmatrix}$,
$\frac{1}{2}\begin{pmatrix}
  1\\ 
  -i\\
  -i\\
   -1
\end{pmatrix}$,
$\frac{1}{2}\begin{pmatrix}
  1\\ 
  i\\
  -i\\
   1
\end{pmatrix}$\\
\hline
  4 &  
$\frac{1}{2}\begin{pmatrix}
  1\\ 
  1\\
  i\\
   -i
\end{pmatrix}$,
$\frac{1}{2}\begin{pmatrix}
  1\\ 
  -1\\
  i \\
   i
\end{pmatrix}$,
$\frac{1}{2}\begin{pmatrix}
  1\\ 
  1\\
  -i\\
   i
\end{pmatrix}$,
$\frac{1}{2}\begin{pmatrix}
  1\\ 
  -1\\
  -i\\
   -i
\end{pmatrix}$\\
\hline
  5 &  
$\frac{1}{2}\begin{pmatrix}
  1\\ 
  -i\\
   1\\
  i
\end{pmatrix}$,
$\frac{1}{2}\begin{pmatrix}
  1\\ 
  i\\
   1\\
   -i
\end{pmatrix}$,
$\frac{1}{2}\begin{pmatrix}
  1\\ 
  -i\\
  -i\\
   -i\\
\end{pmatrix}$,
$\frac{1}{2}\begin{pmatrix}
  1\\ 
  i\\
  -1\\
   i
\end{pmatrix}$\\
\hline
\end{tabular}
\end{center}
\caption{\label{table3} The MUBs associated with lines of the $4 \times 4$ discrete phase space of two-qubit systems.}
\end{table}
The DWFs are defined using the phase space point operators $\textbf{A}_{\alpha}$ as shown in Eq. (\ref{DWFformula}). Depending on the associations of a set of MUB vectors with striations we have $4^{4+1}$ different $\textbf{A}_{\alpha}$'s and thus, $4^{4+1}$ possible DWFs. Out of $4^{4+1}$ possible $\textbf{A}_{\alpha}$'s, 320 have the spectrum ($-0.5000$, $-0.5000$, $0.1339$, $1.866$), other 320 have ($-0.8661$, $-0.5000$, $0.8661$, $1.5000$), and the remaining 384 have the spectrum ($-0.8968$, $-0.1420$, $0.2787$, $1.7601$) as elaborated in \cite{casaccino2008extrema}. By trying out different associations of MUB vectors with striations, we find three $A_{\alpha}$'s, where each $A_{\alpha}$ has a different spectrum. The phase space point operator $A_{\alpha}$ at phase space point $\alpha(1, 1)$ having the the spectrum ($-0.8968$, $-0.1420$, $0.2787$, $1.7601$) is given by
\begin{equation}
   \begin{aligned}
    \textbf{A}_{(1, 1)} = \left(
\begin{array}{cccc}
 0 & -\frac{1}{2}-\frac{i}{2} & \frac{1}{2}-\frac{i}{2} & -\frac{1}{2} \\
 -\frac{1}{2}+\frac{i}{2} & 0 & \frac{i}{2} & 0 \\
 \frac{1}{2}+\frac{i}{2} & -\frac{i}{2} & 1 & 0 \\
 -\frac{1}{2} & 0 & 0 & 0 \\
\end{array}
\right).
      \end{aligned}
      \label{A_NS1}
\end{equation}
The state $NS_1$ given in Eq. (\ref{negative_quantum_states}) is the normalized eigenvector of the above phase space point operator corresponding to the most negative eigenvalue, {\it viz.}, -0.8968. Similarly, the $NS_2$ and the $NS_3$ states can be obtained from the phase space point operators that have the remaining two spectra. 
\bibliographystyle{iopart-num}
\bibliography{BibTexfile}
\end{document}